\documentclass[9pt,shortpaper,twoside,onecolumn,web]{ieeecolor}
\usepackage{generic}
\usepackage{cite}
\usepackage{graphicx}          
\usepackage{ifpdf}
\usepackage{amsmath,amssymb,amsfonts}
\usepackage{mathrsfs}
\usepackage{algorithmic}
\usepackage{algorithm}
\usepackage{graphicx}
\usepackage{textcomp}
\usepackage{color}
\usepackage{enumerate}
\usepackage{epstopdf}
\usepackage{epsfig}
\def\BibTeX{{\rm B\kern-.05em{\sc i\kern-.025em b}\kern-.08em
    T\kern-.1667em\lower.7ex\hbox{E}\kern-.125emX}}
\begin{document}
\title{Replay Attack Detection Based on Parity Space Method for Cyber-Physical Systems}
\author{Dong Zhao, Yang Shi, \IEEEmembership{Fellow, IEEE}, Steven X. Ding, Yueyang Li, Fangzhou Fu
\thanks{Dong Zhao is with the School of Cyber Science and Technology, Beihang University, Beijing 100191, China (e-mail: dzhao@buaa.edu.cn).
        }
\thanks{Yang Shi is with the Department of Mechanical Engineering, University of Victoria, Victoria, BC, V8W 2Y2, Canada (e-mail: yshi@uvic.ca).}
\thanks{Steven X. Ding is with the Institute for Automatic Control and Complex Systems, University of Duisburg-Essen, 47057, Duisburg, Germany (e-mail: steven.ding@uni-due.de).}
\thanks{Yueyang Li is with the School of Electrical Engineering, University of Jinan, Jinan 250022, China (e-mail: cse\_liyy@ujn.edu.cn).}
\thanks{Fangzhou Fu is with the School of Aeronautics and Astronautics, Sun Yat-sen University, Shenzhen 518107, China (e-mail: fufzh@mail.sysu.edu.cn).}
}

\maketitle

\begin{abstract}
The replay attack detection problem is studied from a new perspective based on parity space method in this paper. The proposed detection methods have the ability to distinguish system fault and replay attack, handle both input and output data replay, maintain certain control performance, and can be implemented conveniently and efficiently. First, the replay attack effect on the residual is derived and analyzed. The residual change induced by replay attack is characterized explicitly and the detection performance analysis based on two different test statistics are given. Second, based on the replay attack effect characterization, targeted passive and active design for detection performance enhancement are proposed. Regarding the passive design, four optimization schemes regarding different cost functions are proposed with optimal parity matrix solutions, and the unified solution to the passive optimization schemes is obtained; the active design is enabled by a marginally stable filter so as to enlarge the replay attack effect on the residual for detection. Simulations and comparison studies are given to show the effectiveness of the proposed methods.
\end{abstract}


\section{Introduction}
Cyber-physical systems, which integrate sensing, computation, control, and communication into physical systems, provide us an efficient framework for complex system information service, control, and real-time perception. Increasing applications of cyber physical systems have been demonstrated in recent years, including water distribution systems, power systems, industrial control, and intelligent transportation. As essential parts of cyber physical systems, control systems have been largely energized and also remolded within the cyber-physical framework. The cyber-physical framework strengthens control systems in many directions and at the same time, induces the risk of cyber security and vulnerability of cyber attacks as well, which can cause disastrous consequences demonstrated by living examples \cite{1_ding2018,2_zhou2021}.

From the viewpoint of implementation resources and destructive effects, cyber attacks for control systems are generally classified into disclosure attacks, deception attacks, and disruption attacks \cite{3_dibaji2019,4_teixeira2015}, where eavesdropping attack, zero dynamic attack, replay attack, covert attack, false data injection attack, and denial of service attack are mainly considered \cite{5_sanchez2019}. Specifically, replay attack, which is one of the deception attacks, does not depend on control system model information for implementation and can cause fatal deterioration in control performance while keeping stealthiness (e.g., the Stuxnex \cite{3_dibaji2019}). Thus, driven by the rapidly increasing and strong demands for cyber security against replay attack, a significant amount of research attention has been devoted to replay attack detection.

Since the stealthiness of replay attack builds on the (statistical) feature similarity of control system data in different time periods, the reported detection methods aim to break this similarity or stealthiness for successful replay attack detection. In general, watermarking method, moving target method, and end-to-end encryption method are adopted for replay attack detection. The watermarking method aims to add authentication noise signal to system control signal, and thus the output data replay will break the control signal consistent between the monitoring side and the plant side, which contributes to successful replay attack detection. The basic idea of watermarking method regarding residual distribution property check is presented in \cite{6_mo2009}, the so called dynamic watermarking method is proposed for different control system signal distribution check \cite{7_satchidanandan2017,8_hespanhol2017,9_porter2021,10_satchidanandan2020}, and the periodic watermarking design is given in \cite{11_fang2020}. As the authentication noise signal leads to control performance loss, trade-off between control and replay attack detection performance is studied in \cite{12_weerakkody2014,13_miao2013}. For the moving target method, extra system dynamics, including extra (virtual) control channels, sensing channels, and system states, are added and most importantly, time-varying elements, which are independent of the attacker are introduced to system dynamics (e.g., gain switching or time-varying parameters). In this case, replay attack can be detected since the output data similarity for the detector along time does not hold. Excellent results of moving target method, including replay attack detection performance analysis and optimization, can be found in \cite{14_griffioen2019,15_weerakkody2015,16_griffioen2021} and references therein. The end-to-end encryption method is the most adopted method for ensuring data security in communication based on encryption algorithms, where the secret key of encryption builds the security of data as well as the possibility of anomaly detection. For cyber-physical control systems, various realizations of the classical end-to-end encryption idea have been exploited. To mention a few, output filter based encryption and decryption scheme is proposed for attack detection in \cite{17_ferrari2021} and characterized residual signal generation as well as gain switching for encrypted signal transmission is proposed in \cite{18_ding2022} for general deception attack detection. Direct sensor signal coding by random noise for attack detection is proposed in \cite{19_ye2019}. The spectral estimation based replay attack detection method with communication error related hidden encryption for output data is given in \cite{20_tang2015}. The generalized control signal encryption and decryption scheme realized by disturbance compensation is given in \cite{21_trapiello2019}. The input signal based output signal encryption is studied in \cite{22_guo2021}.

The reported replay attack detection methods are powerful, supported by different implementation resources and accompanied by different levels of control performance cost. However, the following issues need a further consideration for replay attack detection:
\begin{enumerate}
  \item it is much more challenging to detect the replay attack with both input and output data replay, especially when the attacker is aware of the existence of the detector;
  \item watermarking input data requires a control performance sacrifice at any time for replay attack detection, even if the control performance is significant for the user;
  \item it is hard for the reported detection methods to distinguish system fault and replay attack by the detector, where the user may be confused and the attack warning is ignored even if an alarm is triggered by the replay attack detector;
  \item the moving target method requires certain plant-side implementation resources due to the deep coupling between the original system dynamics and the added time-varying dynamics.
\end{enumerate}

Motivated by the discussed issues above, we come back to the replay attack effect on control system directly and try to exploit the intrinsic behavior of replay attack as well as the attacked control system dynamics for detection. In this paper, we study and propose a new scheme for replay attack detection based on parity space method. The proposed scheme has the following distinctive features: (1) maintaining the control performance; (2) handling both input and output data replay attack; (3) isolating replay attack from system faults; (4) not requiring the real-time secret key synchronization; and (5) easy implementation in both cyber and physical domains.

\begin{figure}
\begin{center}
\epsfig{file=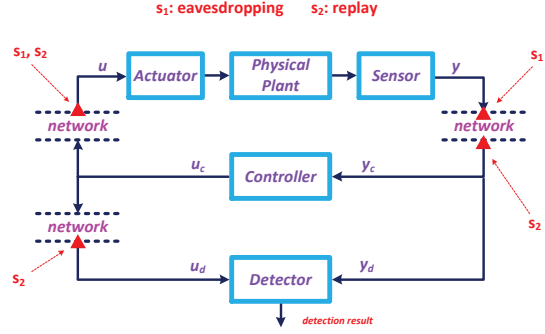,width=7.448805cm,height=4.51269cm}
\caption{Closed-loop control and monitoring configuration of cyber-physical systems under replay attack.}
\label{fig1}
\end{center}
\end{figure}

The classical parity space method has been well studied for fault diagnosis \cite{23_ding2008}, but the very important ideas for adopting the parity space method here are its finite horizon property and real-time input-output data parity check, e.g., forming a system kernel space \cite{18_ding2022}. By employing the parity space based residual generation, the replay attack detection problem is solved in three steps: catch the replay attack effect, analyze the replay attack effect, and enlarge the replay attack effect. Specifically, the feasibility and advantages of using the parity space method for replay attack detection are analyzed, the replay attack effect on the residual by means of the parity space method is quantified explicitly, and the associated replay attack detection conditions and performance are addressed quantitatively. Thus, the security level of the considered cyber-physical system can be known. After the replay attack effect catch and analysis, targeted detection methods based on parity space formulation are proposed to improve the detection performance by enlarging the replay attack effect. Both passive and active detector design methods are proposed. About the passive detection method, four optimal parity matrices are derived corresponding to different optimization indices, and the relationship among these parity matrices are discussed for a unified solution. For the active detector design method, based on the amplitude and occurrence analysis of the replay attack induced residual change, a marginally stable filter is introduced to the control system to improve the detection performance further.

This paper is organized as follows. The system description and problem formulation are given in Section II. Replay attack effect quantification and detection performance analysis are given in Section III. Detection performance enhancement with passive and active design is given in Section IV. The proposed detection methods are illustrated by simulation study in Section V, and Section VI provides some concluding remarks.

\textbf{Notation}. For vectors $a$ and $b$, $col\{ {a},{b}\}  = {\left[ {\begin{array}{*{20}{c}}{a^T}&{b^T}\end{array}} \right]^T}$, $\mathcal{E}\{a\}$ denotes the expectation of $a$, $cov\{a\}$ denotes the covariance (matrix) of $a$, and $cov\{a,b\}$ denotes the covariance (matrix) between $a$ and $b$. For a square matrix $X$, $det(X)$ and $tr(X)$ are the determinant and trace of $X$, respectively; $X>0$ ($X<0$) denotes a positive (negative) definite matrix $X$. $\left\|X\right\|_i$ is the $i$-norm of $X$ ($i$ is omitted when $i=2$). $X^T$, $X^{-1}$, $X^{\bot}$, and $X^+$ are the transpose, inverse, orthogonal complement, and Moore-Penrose inverse of $X$, respectively. $diag\{X_1,X_2\}$ is the block diagnal matrix formed with $X_1$ and $X_2$. $N(\mu,\Sigma)$ denotes the Gaussian distribution with mean $\mu$ and covariance $\Sigma$. $*$ is used for denoting symmetric blocks in matrices. For simplicity, we omit the variables in functions if it is clear from the context.

\section{Preliminaries}
In this section, the considered cyber-physical system configuration, replay attack description, and parity space method are introduced. Then, the replay attack detection problem is formulated.
\subsection{Cyber-Physical System Configuration}
Consider the following linear time-invariant system:
\begin{eqnarray}\label{1}
\left\{ \begin{array}{l}
x(k + 1) = Ax(k) + {B_u}u(k) + {B_w}w(k)\\
y(k) = Cx(k) + v(k)
\end{array} \right.
\end{eqnarray}
where $x(k) \in \mathcal{R}^n$ is the system state, $u(k) \in \mathcal{R}^m$ is the system input, and $y(k) \in \mathcal{R}^p$ is the measurement output. $w(k) \in \mathcal{R}^q$ and $v(k)$ represent the process and measurement noises, respectively. Without loss of generality, we assume that $w(k)$ and $v(k)$ are zero mean Gaussian white noises satisfying
\begin{displaymath}
{cov} \{ {\left[ {\begin{array}{*{20}{c}}
{{w^T}(k)}&{{v^T}(k)}\end{array}} \right]^T}\}  = diag\{ Q,R\},
\end{displaymath}
where $Q >0$ and $R>0$ are known matrices. $A, B_u, B_w$, and $C$ are known matrices with appropriate dimensions.

The control and monitoring configuration of the cyber-physical system is shown in Fig. 1, where
\begin{itemize}
  \item $y$ is the system output, $y_c$ is the received system output for the controller, $y_d$ is the received system output for the detector,
  \item $u$ is the system input, $u_c$ is the controller output, and $u_d$ is the received control signal for the detector.
\end{itemize}
The detector is employed to detect the potential anomaly (attacks or faults), and the controller is used to ensure a certain control performance of the system, e.g., stability or tracking performance. The controller is assumed to be linear, either static or dynamic, e.g., linear quadratic Gaussian (LQG) controller. The control and measurement signals are transmitted via the network. Moreover, due to the cyber-physical system framework, the control system may encounter cyber attacks.

\subsection{Replay Attack Description}
In this section, the description of replay attack associated with the considered cyber-physical system is given.

Without any attack, the signal transmission between the cyber part and the physical part is denoted by
\begin{eqnarray}\label{2}
\left\{ \begin{array}{l}
y(k) = {y_c}(k) = {y_d}(k),\\
{u}(k) = u_c(k) = {u_d}(k).
\end{array} \right.
\end{eqnarray}

For implementation of replay attack, the attackers have the ability to
\begin{itemize}
  \item read any sensor output and/or control input transmissions, and
  \item modify any sensor output and/or control input transmissions.
\end{itemize}
To perform a stealthy replay attack, a two-stage action is implemented.
\begin{description}
  \item[\emph{stage-$1$}]~Eavesdropping: record the accessible sensor output and/or control input transmissions for a sufficiently long time period $[{t_{r1}},{t_{r2}}]$, where ${t_{r1}}, {t_{r2}} \in {\mathcal{N}^ + }$ and \eqref{2} holds.
  \item[\emph{stage-$2$}]~Replay: replace sensor output transmissions since $k\ge T_0$ by the previously recorded data (${T_0} \gg {t_{r2}}$), where ${y_c}(k) = {y_d}(k) = y({\tau _k})$, ${t_{r1}} \le {\tau _k} \ll {t_{r2}}$, and ${T_0} \in {\mathcal{N}^ + }$; add a malicious signal ${u_a}(k)$ to the system control input: $u(k) = {u_c}(k) + {u_a}(k)$; and/or replace the control input transmissions for the detector by the previously recorded data such that ${u_d}(k) = u({\tau _k})$, otherwise, ${u_d}(k) = {u_c}(k)$.
\end{description}

To keep the generality of the replay attack detection problem, the following Assumption is given.

\textbf{Assumption 1}. \emph{In the absence of replay attack, $\mathcal{E}\{y(k)\}$ and $cov\{y(k)\}$ are constant.}

Assumption 1 can be achieved when system \eqref{1} has run in steady state for a sufficient long time period. The generality of Assumption 1 for replay attack detection will be explained later (See Section III.A).

Thus, under Assumption 1, the replay attack has the following properties.
\begin{itemize}
  \item The attacker can destroy the control system performance based on the signal $u_a(k)$, while the data replacement of ${y_c}(k)$ opens the control loop;
  \item The attacker can remain stealthy by data fraud, i.e., $y({\tau _k}) \approx y(k)$ and $u({\tau _k}) \approx u(k)$ when $k \ge {T_0}$.
\end{itemize}

As $u_a(k)$ is not accessible for the detector, $u_d(k)$ and $y_d(k)$ is the available source data for replay attack detection, which ensures the basic stealthiness of replay attack.

\subsection{Parity Space Method}
For anomaly detection purpose, the residual generation based on parity space method is introduced in this section.

Define
\begin{align}
{Y_s}(k) &= col\left\{ {y(k - s),y(k - s + 1), \cdots ,y(k)} \right\},\nonumber\\
{U_s}(k) &= col\left\{ {u(k - s),u(k - s + 1), \cdots ,u(k)} \right\},\nonumber\\
{W_s}(k) &= col\left\{ {w(k - s),w(k - s + 1), \cdots ,w(k)} \right\},\nonumber\\
{V_s}(k) &= col\left\{ {v(k - s),v(k - s + 1), \cdots ,v(k)} \right\}.\nonumber
\end{align}
The parity relation based on \eqref{1} is given by
\begin{align}
{Y_s}(k) = {H_{0,s}}x(k - s) + {H_{u,s}}{U_s}(k) + {E_s}(k),\label{3}
\end{align}
where ${E_s}(k) = {H_{w,s}}{W_s}(k) + {H_{v,s}}V(k)$, $s$ is the order of parity relation, and
\begin{displaymath}
\begin{array}{l}
\begin{aligned}
{H_{u,s}} &= \left[ {\begin{array}{*{20}{c}}
0&0& \cdots &0\\
{C{B_u}}&0& \ddots & \vdots \\
 \vdots & \ddots & \ddots &0\\
{C{A^{s - 1}}{B_u}}& \cdots &{C{B_u}}&0
\end{array}} \right],\\
{H_{w,s}} &= \left[ {\begin{array}{*{20}{c}}
0&0& \cdots &0\\
{C{B_w}}&0& \ddots & \vdots \\
 \vdots & \ddots & \ddots &0\\
{C{A^{s - 1}}{B_w}}& \cdots &{C{B_w}}&0
\end{array}} \right],\\
{H_{v,s}} &= \left[ {\begin{array}{*{20}{c}}
{{I_p}}&0& \cdots &0\\
0&{{I_p}}& \ddots & \vdots \\
 \vdots & \ddots & \ddots &0\\
0& \cdots &0&{{I_p}}
\end{array}} \right],~{H_{0,s}} = \left[ {\begin{array}{*{20}{c}}
C\\
{CA}\\
 \vdots \\
{C{A^s}}
\end{array}} \right].
\end{aligned}
\end{array}
\end{displaymath}
For parity relation check \cite{23_ding2008}, the residual $r_s(k)$ based on the accessible input and output data is given by
\begin{align}
{r_s}(k) = {Z_s}\left( {{Y_s}(k) - {H_{u,s}}{U_s}(k)} \right),\label{4}
\end{align}
where ${Z_s} \in {\mathcal{R} ^{l \times p(s + 1)}}$ is the parity matrix. Define ${z_{s,i}}$ the $i$-th row of $Z_s$ which is delivered from its parity space ${P_s} = H_{0,s}^ \bot $.

Based on \eqref{3} and the definition of $Z_s$, we have
\begin{align}
{r_s}(k) = {Z_s}{E_s}(k) \sim N(0,{\Theta _s}),\label{5}
\end{align}
where
\begin{displaymath}
\begin{array}{l}
{\Theta _s} = {Z_s}{H_{w,s}}{Q_s}H_{w,s}^TZ_s^T + {Z_s}{H_{v,s}}{R_s}H_{v,s}^TZ_s^T.\\
{Q_s} = {I_{s + 1}} \oplus Q,~{R_s} = {I_{s + 1}} \oplus R.
\end{array}
\end{displaymath}
Based on \eqref{4} and \eqref{5}, an anomaly detector can be designed based on the mean and covariance change detection of the residual $r_s(k)$. In general, $\chi^2$ and generalized likelihood ratio (GLR) based test statistics can be used (the motivation and applicability of these two test statistics will be given in Section III).

\subsection{Problem Formulation}
Recalling the implementation property of replay attack, a switching in data for dynamic systems is necessary, where this switching will cause data ``ripples" even if system \eqref{1} is in steady state. Thus, we try to catch and quantify the data ``ripples" triggered by this switching for detection. Just like the natural scene that water ``ripples" will vanish after a long time period since its occurrence, the data ``ripples" induced by replay attack will converge to zero in stable closed-loop system, which means that timely detection of replay attack is essential. This enlightens us to focus on finite horizon detection schemes with respect to system dynamics, where the parity space method is adopted. Besides, the control performance degradation, fault and replay attack isolation, and simultaneous input and output replay are all open issues to be addressed for replay attack detection.

Based on the above analysis, the detection problem of replay attack using parity space method for system \eqref{1} is studied to solve these open issues. Generally, under all these claimed ``expectations" above for replay attack detection, two major questions are to be answered: whether the parity space method works for replay attack detection and how to improve the detection performance if the parity space method works. Specifically, the key problems are to
\begin{enumerate}
  \item catch and quantify the residual change driven by the replay attack,
  \item quantify the performance of replay attack detection based on parity space method,
  \item design the optimal weighting matrix regarding the residual change for performance enhancement of replay attack detection,
  \item and reconfigure the detection scheme to enhance detection performance in the framework of parity space formulation.
\end{enumerate}
The cases with and without input data replay are comprehensively considered. It is worthwhile to highlight that by applying the proposed new scheme, the control performance sacrifice is not needed here, and the proposed detection scheme can be implemented conveniently and efficiently and distinguish physical fault and replay attack.

\textbf{Remark 1}. \emph{Both passive and active detection/design methods are introduced in this study. The passive detection method means that no extra signal or system dynamics is added to the plant side for detection. Different from the passive detection method, extra excitation/encryption signal and/or dynamics is used for active detection, e.g., the watermarking method \cite{6_mo2009}, encryption method, and moving target method \cite{3_dibaji2019}. The key point for active detection is the ratio between the performance improvement and the implementation cost of the method. Generally, the passive method and active method feature their own advantages for implementation and detection performance, respectively. In this study, we propose satisfying passive detection method and also active scheme for diverse applications.}

\section{Qualitative and Quantitative Performance Analysis of Parity Space Method for Replay Attack Detection}
To achieve successful replay attack detection, the key is to catch and then amplify the attack effect on the residual. We start with analyzing the statistical changes in residual due to replay attack, where the observed changes lead to the feasibility/condition for replay attack detection via parity space method. Both qualitative and quantitative detection performances are analyzed, and the cases with and without input data replay are all considered.

\subsection{Quantification of the Replay Attack Effect}
Observing \eqref{3} and \eqref{5}, from the viewpoint of the detector, the system input and output form a perfect matching in the absence of any attack (and fault), or the input and output data belong to the kernel space of the system \cite{23_ding2008}. However, what will happen to the equation in \eqref{3} in the presence of replay attack? To answer this question, we will show the statistical changes triggered by the replay attack on \eqref{3} as well as $r_s(k)$.

For $k\ge T_0$, the replay attack is in stage-$2$. Thus, $y_d(k)$ and $u_d(k)$ are replaced by $y(\tau_k)$ and $u(\tau_k)$, respectively (Note: the case without input data replay will be discussed later). Let $\alpha \in \mathcal{N}$ be the step number of data replacement, i.e., $\alpha  = k - {T_0} + 1$. Thus, $\alpha =0$ means attack-free, $\alpha  > s$ means full data replacement for parity relation, and $0 < \alpha  \le s$ means partial data replacement for parity relation.

In the following, we will analyze the change of residual due to replay attack.

\subsubsection{Residual Changes with Full Data Replay}
Define
\begin{displaymath}
\begin{array}{l}
\begin{aligned}
{Y_{d,s}}(k) &= col\left\{ {{y_d}(k - s),{y_d}(k - s + 1), \cdots ,{y_d}(k)} \right\},\\
{U_{d,s}}(k) &= col\left\{ {{u_d}(k - s),{u_d}(k - s + 1), \cdots ,{u_d}(k)} \right\},\\
{U_{c,s}}(k) &= col\left\{ {{u_c}(k - s),{u_c}(k - s + 1), \cdots ,{u_c}(k)} \right\}.
\end{aligned}
\end{array}
\end{displaymath}
When $\alpha >s$, it is known from the replay attack description that ${Y_{d,s}}(k) = {Y_s}({\tau _k})$, and thus
\begin{align}
{Y_{d,s}}(k) =&~ {H_{0,s}}x({\tau _k} - s) + {H_{u,s}}{U_{d,s}}(k) \nonumber\\
&+ {H_{u,s}}[{U_s}({\tau _k}) - {U_{d,s}}(k)] + {E_s}({\tau _k})\label{6}.
\end{align}
When the input data encounters replay attack, we have ${U_s}({\tau _k}) = {U_{d,s}}(k)$, so the residual is given by
\begin{align}
r_s^\alpha (k) &= {Z_s}\left( {{Y_{d,s}}(k) - {H_{u,s}}{U_{d,s}}(k)} \right)\nonumber\\
 &= {Z_s}\left( {{Y_s}({\tau _k}) - {H_{u,s}}{U_s}({\tau _k})} \right)\nonumber\\
 &= {Z_s}{E_s}({\tau _k})={r_s}({\tau _k}).\label{7}
\end{align}
It is obvious that the residual signal $r_s^\alpha (k)$ has the same statistical characteristics as ${r_s}({\tau _k})$ under both input and output data replay for the detector when $\alpha >s$. However, with only output data replay, one has ${U_{c,s}}(k) = {U_{d,s}}(k)$ and
\begin{align}
r_s^\alpha (k) &= {Z_s}\left( {{Y_{d,s}}(k) - {H_{u,s}}{U_{d,s}}(k)} \right)\nonumber\\
 &= {Z_s}\left( {{Y_s}({\tau _k}) - {H_{u,s}}{U_{c,s}}(k)} \right)\nonumber\\
 &= {r_s}({\tau _k}) + {Z_s}{H_{u,s}}[{U_s}({\tau _k}) - {U_{c,s}}(k)].\label{8}
\end{align}

Now, the change of residual induced by only output data replay lies in ${U_s}({\tau _k}) - {U_{c,s}}(k)$ when $\alpha  > s$ . If a static controller is adopted, ${U_s}({\tau _k}) - {U_{c,s}}(k) = 0$ when $\alpha >s$ and $k>T_0$, because the same static controller outputs the same control signals at two different time instants with the same input vectors ${Y_s}({\tau _k})$. If a linear stable dynamic controller is adopted, $\mathop {\lim }\limits_{\alpha  \to \infty } [{U_s}({\tau _k}) - {U_{c,s}}(k)] = 0$ for $k \gg T_0$, where the influence of the controller initial value discrepancy at two different time instants will vanish due to the convergence property of the controller when $\alpha  \to \infty $. All in all, the above analysis together shows that replay of both input and output data will largely increase the stealthiness of replay attack, since ${U_s}({\tau _k}) - {U_{d,s}}(k) = 0$ holds in spite of the controller structure during the time period of input data replay.

In practice, under the convergence property of the controllers, $\left\| {u({\tau _k}) - u(k)} \right\|$ will be relatively small when $\alpha >s$ and $k>T_0$. Thus, it is more promising to detect the replay attack when $0<\alpha \le s$, and thus the change of the statistical property for $r_s^\alpha (k)$ due to the controller initial value discrepancy is analyzed in the next subsection.

\textbf{Remark 2}. \emph{The well-known watermarking technique for control input works for replay attack detection, based on the statistical characteristics verification of residual at each time instant. However, the watermarking technique may fail when input data replay is activated for detector based on the above analysis (based on \eqref{7} when $\alpha >s$, the statistical characteristic of residual is not distinguishable from the attack-free scenario).}

\subsubsection{Residual Changes with Partial Data Replay}
With both input and output data replay for the detector, Equation \eqref{3} is rewritten as
\begin{align}
Y_s^\alpha (k,{\tau _k}) = &~{H_{0,s}}x(k - s) + {H_{u,s}}U_s^\alpha (k,{\tau _k})\nonumber\\
 &+ {E_s}(k) + \Gamma _s^\alpha (k),\label{9}
\end{align}
where
\begin{displaymath}
Y_s^\alpha (k,{\tau _k}) = \left\{ \begin{array}{l}
\begin{aligned}
&{Y_s}(k),&\alpha  = 0\\
&{Y_s}({\tau _k}),&\alpha  > s\\
&\left[ {\begin{array}{*{20}{c}}
{{Y_{s - \alpha }}(k - \alpha )}\\
{{Y_{\alpha  - 1}}({\tau _k})}
\end{array}} \right],&else
\end{aligned}
\end{array} \right.
\end{displaymath}
\begin{displaymath}
\begin{array}{l}
\begin{aligned}
\Gamma _s^\alpha (k) =~& H_{0,s}^\alpha (x({\tau _k} - s) - x(k - s))\\
 &+ H_{u,s}^\alpha \left[ {\begin{array}{*{20}{c}}
{{U_{s - \alpha }}({\tau _k} - \alpha ) - {U_{s - \alpha }}(k - \alpha )}\\
0
\end{array}} \right]\\
 &+ H_{w,s}^\alpha ({W_s}({\tau _k}) - {W_s}(k)) + H_{v,s}^\alpha ({V_s}({\tau _k}) - {V_s}(k)),
\end{aligned}
\end{array}
\end{displaymath}
\begin{displaymath}
H_{0,s}^\alpha  = \left\{ \begin{array}{l}
\begin{aligned}
&0,&\alpha  = 0\\
&{H_{0,s}},&\alpha  > s\\
&{H_{0,s}} - \left[ {\begin{array}{*{20}{c}}
{{H_{0,s - \alpha }}}\\
{{0_{p\alpha  \times n}}}
\end{array}} \right],&else,
\end{aligned}
\end{array} \right.
\end{displaymath}

\begin{displaymath}
H_{u,s}^\alpha  = \left\{ \begin{array}{l}
0,~\alpha  = 0\\
{H_{u,s}},~\alpha  > s\\
{H_{u,s}} - \left[ {\begin{array}{*{20}{c}}
{{H_{u,s - \alpha }}}&{{0_{p(s + 1 - \alpha ) \times \alpha m}}}\\
{{0_{p\alpha  \times (s + 1 - \alpha )m}}}&{{H_{u,\alpha  - 1}}}
\end{array}} \right],~else,
\end{array} \right.
\end{displaymath}

\begin{displaymath}
H_{w,s}^\alpha  = \left\{ \begin{array}{l}
0,\alpha  = 0\\
{H_{w,s}},\alpha  > s\\
{H_{w,s}} - \left[ {\begin{array}{*{20}{c}}
{{H_{w,s - \alpha }}}&{{0_{p(s + 1 - \alpha ) \times \alpha q}}}\\
{{0_{p\alpha  \times (s + 1 - \alpha )q}}}&{{0_{p\alpha  \times \alpha q}}}
\end{array}} \right],else,
\end{array} \right.
\end{displaymath}

\begin{displaymath}
H_{v,s}^\alpha  = \left\{ \begin{array}{l}
0,\alpha  = 0\\
{H_{v,s}},\alpha  > s\\
\left[ {\begin{array}{*{20}{c}}
{{0_{p(s + 1 - \alpha ) \times p(s + 1 - \alpha )}}}&*\\
{{0_{p\alpha  \times (s + 1 - \alpha )p}}}&{{I_{p\alpha }}}
\end{array}} \right],else.
\end{array} \right.
\end{displaymath}
and $U_s^\alpha (k,{\tau _k})$ can be obtained by submitting $y$ in $Y_s^\alpha (k,{\tau _k})$ with $u$.

Based on \eqref{9}, the parity space based residual can be generated as
\begin{align}
r_s^\alpha (k) = {Z_s}\left( {Y_s^\alpha (k,{\tau _k}) - {H_{u,s}}U_s^\alpha (k,{\tau _k})} \right).\label{10}
\end{align}

The equation in \eqref{9} denotes the mathematical relation between the accessible system input and output in the presence of replay attack. As the process noise, measurement noise, and state information are not precisely known, only the statistical property of $r_s^\alpha (k)$ for replay attack detection can be checked. Specifically, since the general detection logic and threshold are designed based on attack-free residual $r_s(k)$, we will discuss the statistical difference between $r_s(k)$ and $r_s^\alpha (k)$ resulted by replay attack. Both the mean and covariance changes will be addressed. The main results are given in the following propositions.

\textbf{Proposition 1 (mean change of the residual)}.  \emph{Under Assumption 1 and both input and output data replay for the detector, it holds that
\begin{align}
{\mathcal{E}}\{r_s^\alpha (k) \}= 0, \label{11}
\end{align}
when $0<\alpha \le s$.}

\begin{proof}
Under both input and output data replay, it is known that ${Y_{d,s}}(k) = Y_s^\alpha (k,{\tau _k})$ and ${U_{d,s}}(k) = U_s^\alpha (k,{\tau _k})$. When $0 < \alpha  \le s$, by taking \eqref{9} into \eqref{10}, it yields
\begin{align}
r_s^\alpha (k) &= {Z_s}\left( {Y_s^\alpha (k,{\tau _k}) - {H_{u,s}}U_s^\alpha (k,{\tau _k})} \right)\nonumber\\
 &= {Z_s}\left( {{Y_s}(k) - {H_{u,s}}{U_s}(k)} \right) + {Z_s}\Gamma _s^\alpha (k)\nonumber\\
 &= {Z_s}{E_s}(k) + {Z_s}\Gamma _s^\alpha (k).\label{12}
\end{align}
Following from \eqref{12}, it has
\begin{align}
{\mathcal{E}}\{r_s^\alpha (k)\} = {\mathcal{E}}\{{Z_s}\Gamma _s^\alpha (k)\}.\label{13}
\end{align}
Reformulating $\Gamma _s^\alpha (k)$ in \eqref{9} into a compact form, it yields
\begin{align}
\Gamma _s^\alpha (k) = &\left[ {\begin{array}{*{20}{c}}
{{0_{s + 1 - \alpha }}}\\
{{Y_{\alpha  - 1}}({\tau _k}) - {Y_{\alpha  - 1}}(k)}
\end{array}} \right] \nonumber\\
&- \left[ {\begin{array}{*{20}{c}}
{{0_{s + 1 - \alpha }}}\\
{{H_{u,\alpha  - 1}}{U_{\alpha  - 1}}({\tau _k}) - {H_{u,\alpha  - 1}}{U_{\alpha  - 1}}(k)}
\end{array}} \right].\label{14}
\end{align}
Based on Assumption 1, $\mathcal{E}\{y(i)\}=\mathcal{E}\{y(j)\}$, where $i,j \in \mathcal{N}^+$, it is known that
\begin{align}
{\mathcal{E}}\{{Y_{\alpha  - 1}}({\tau _k})\} = {\mathcal{E}}\{{Y_{\alpha  - 1}}(k)\},\label{15}
\end{align}
\begin{align}
{\mathcal{E}}\{{U_{\alpha  - 1}}({\tau _k})\} = {\mathcal{E}}\{{U_{\alpha  - 1}}(k)\}.\label{16}
\end{align}
Finally, \eqref{11} is proved by taking \eqref{14}, \eqref{15}, and \eqref{16} into \eqref{13}.
\end{proof}

Proposition 1 reveals the relative mean change of the residual in finite horizon due to replay attack under Assumption 1. Based on \eqref{11}, it holds that ${\mathcal{E}}\{r_s^\alpha (k)\} = {\mathcal{E}}\{{r_s}(k)\}$ as $w(k)$ and $v(k)$ are weak stationary processes. If Assumption 1 does not hold due to the transient behavior (initial value effect or reference change), ${\mathcal{E}}\{y(i)\} \ne {\mathcal{E}}\{y(j)\}$ when $i \ne j$, which leads to ${\mathcal{E}}\{r_s^\alpha (k)\} \ne 0$; in such a case, the replay attack can be detected easily. Thus, the attacker usually avoids implementing the data replay when ${\mathcal{E}}\{y(i)\} \ne {\mathcal{E}}\{y(j)\}$. This is the direct reason for introducing Assumption 1 to ensure the generality of this study, from the aspects of both the defender and attacker.

\textbf{Remark 3}. \emph{Even if ${\mathcal{E}}\{r_s^\alpha (k)\} = {\mathcal{E}}\{{Z_s}\Gamma _s^\alpha (k)\} = 0$ is proved, it is of interest to exclude the case that $\Gamma _s^\alpha (k) = 0$. Since
\begin{align}
\Gamma _s^\alpha (k) =& \left[ {\begin{array}{*{20}{c}}
{{0_{s + 1 - \alpha }}}\\
{{Y_{\alpha  - 1}}({\tau _k}) - {H_{u,\alpha  - 1}}{U_{\alpha  - 1}}({\tau _k})}
\end{array}} \right] \nonumber\\
&- \left[ {\begin{array}{*{20}{c}}
{{0_{s + 1 - \alpha }}}\\
{{Y_{\alpha  - 1}}(k) - {H_{u,\alpha  - 1}}{U_{\alpha  - 1}}(k)}
\end{array}} \right],\label{17}
\end{align}
it is almost sure that
\begin{displaymath}
{Y_{\alpha  - 1}}({\tau _k}) - {H_{u,\alpha  - 1}}{U_{\alpha  - 1}}({\tau _k}) \ne {Y_{\alpha  - 1}}(k) - {H_{u,\alpha  - 1}}{U_{\alpha  - 1}}(k),
\end{displaymath}
due to the existence of nonzero system noise and the parity relation in \eqref{3}, which means the non-identical residuals for different time instants. Then, it is almost sure that $\Gamma _s^\alpha (k) \ne 0$. The non-zero $\Gamma _s^\alpha (k)$ when $0 < \alpha  \le s$ means the existence of data ``ripple" due to the replay attack as discussed in Section II.}

Regarding the covariance of $r_s^\alpha (k)$, when $0 < \alpha  \le s$, it is known from \eqref{12} that
\begin{align}
{cov} \{ r_s^\alpha (k)\}  = {cov} \{ {r_s}(k) + {Z_s}\Gamma _s^\alpha (k)\}  = \Theta _s^\alpha ,\label{18}
\end{align}
where $\Theta _s^\alpha  = {\Theta _s} + {\Delta _\alpha }$ and
\begin{align}
{\Delta _\alpha } =~& {cov} \{ {Z_s}\Gamma _s^\alpha (k)\}  + {cov} \{ {r_s}(k),{Z_s}\Gamma _s^\alpha (k)\} \nonumber\\
& + {cov} \{ {Z_s}\Gamma _s^\alpha (k),{r_s}(k)\} .\nonumber
\end{align}

Regarding the covariance of $r_s^\alpha (k)$, the following proposition is given.

\textbf{Proposition 2 (covariance change of the residual)}. \emph{Under Assumption 1 and both input and output data replay for the detector, it holds that
\begin{align}
{cov} \{ r_s^\alpha (k)\}  \ne {cov} \{ {r_s}(k)\} ,\label{19}
\end{align}
when $0 < \alpha  \le s$.}

\begin{proof}
Based on \eqref{12}, one knows that
\begin{align}
{cov} \{ r_s^\alpha (k)\}  &= {cov} \{ {Z_s}{E_s}(k) + {Z_s}\Gamma _s^\alpha (k)\} \nonumber\\
 &= {Z_s}{cov} \{ {E_s}(k) + \Gamma _s^\alpha (k)\} Z_s^T.\label{20}
\end{align}
When $0 < \alpha  \le s$, it is known from Remark 3 that $\Gamma _s^\alpha (k) \ne 0$. Moreover, ${Z_s}\Gamma _s^\alpha (k) \ne 0$ and
\begin{displaymath}
\Gamma _s^\alpha (k) \ne  - 2{E_s}(k)
\end{displaymath}
can be concluded by referring to the structure of $\Gamma _s^\alpha (k)$ in \eqref{17}. Recalling the definition ${cov} \{ {r_s}(k)\}  = {Z_s}{cov} \{ {E_s}(k)\} Z_s^T$, \eqref{19} is proved.
\end{proof}

Proposition 2 proves the existence of a nonzero covariance change of the residual due to the replay attack. Thus, Propositions 1 and 2 together depict the replay attack effect on the parity space based residual.

The change of residual caused by replay attack with both input and output data replay is analyzed. It is also of interest to address the case with only output data replay.

Following from the analysis in Section III.A.1 and in the presence of system output data replay, the adopted data for residual generation is
\begin{align}
{Y_d}(k) = Y_s^\alpha (k,{\tau _k}),\label{21}
\end{align}
\begin{align}
{U_d}(k) = U_s^\alpha (k,{\tau _k}) + \left[ {\begin{array}{*{20}{c}}
0\\
{{U_{c,\alpha }}(k) - {U_{c,\alpha }}({\tau _k})}
\end{array}} \right],\label{22}
\end{align}
when $0<\alpha \le s$.

Comparing the adopted data set for residual generation in the cases with and without input data replay, one modification term ${U_{c,\alpha }}(k) - {U_{c,\alpha }}({\tau _k})$ to $U_s^\alpha (k,{\tau _k})$ is needed here for the case with only output data replay. Note that ${U_{c,\alpha }}(k)$ is driven by ${Y_\alpha }({\tau _k})$, which is the input for ${U_{c,\alpha }}({\tau _k})$ as well, due to the system output data replay. ${U_{c,\alpha }}(k) - {U_{c,\alpha }}({\tau _k})$ is the direct result of initial value discrepancy of the dynamic controller as discussed in Section III.A.1.

To characterize ${U_{c,\alpha }}(k) - {U_{c,\alpha }}({\tau _k})$ and without loss of generality (if static controller is considered, ${U_{c,\alpha }}(k) - {U_{c,\alpha }}({\tau _k}) = 0$), a linear dynamic output feedback controller is considered. Assume the controller owns a realization of $\left\{ {{A_c},{B_c},{C_c},{D_c}} \right\}$ with controller state $\hat x(k)$.

Based on \eqref{21} and \eqref{22}, the residual for system output data replay is generated as
\begin{align}
r_s^\alpha (k) = {Z_s}\left( {Y_s^\alpha (k,{\tau _k}) - {H_{u,s}}U_s^\alpha (k,{\tau _k})} \right) + {Z_s}\Gamma _{s,u}^\alpha (k),\label{23}
\end{align}
where
\begin{displaymath}
\Gamma _{s,u}^\alpha (k) = \left[ {\begin{array}{*{20}{c}}
0\\
{{H_{u,\alpha  - 1}}{H_{c,\alpha  - 1}}\left( {\hat x({\tau _k} - \alpha  + 1) - \hat x(k - \alpha  + 1)} \right)}
\end{array}} \right]
\end{displaymath}
and ${H_{c,\alpha  - 1}}$ can be constructed by replace $A$ and $C$ in ${H_{0,\alpha  - 1}}$ by $A_c$ and $C_c$, respectively. Taking \eqref{12} into \eqref{23} yields
\begin{align}
r_s^\alpha (k) = {Z_s}{E_s}(k) + {Z_s}\Gamma _s^\alpha (k) + {Z_s}\Gamma _{s,u}^\alpha (k).\label{24}
\end{align}

Following the proofs for Propositions 1 and 2, the results in \eqref{11} and \eqref{19} cover $r_s^\alpha (k)$ in \eqref{24} as well under Assumption 1. Thus, with or without input data replay, the replay attack will cause a covariance change on the residual comparing to the attack-free case. Based on this conclusion, in the following, the residual covariance change due to replay attack will be further quantified for detection.

\subsection{Quantification of the Residual Covariance under Replay Attack}
Even though replay attack effect does exist on the residual generated by the parity space based detector, quantitative characterization of this effect is essential for successful replay attack detection and specifically, for the targeted detection scheme design. In this Section, the replay attack effect, i.e., the covariance change of residual, is further quantified.

As ${\Delta _\alpha }$ in \eqref{18} is the key change caused by replay attack, we will analyze ${\Delta _\alpha }$ for the case under both system input and output data replay and then extend the result to the case with only system output data replay.

\subsubsection{Residual Covariance Change under System Input and Output Replay}
Since ${\tau _k} \ll {T_0} \le k$, we have
\begin{displaymath}
{cov} \{ {E_s}(k),{Y_{\alpha  - 1}}({\tau _k})\}  = 0,
\end{displaymath}
\begin{displaymath}
{cov} \{ {E_s}(k),{U_{\alpha  - 1}}({\tau _k})\}  = 0,
\end{displaymath}
\begin{displaymath}
{cov} \{ x(k),x({\tau _k})\}  = 0.
\end{displaymath}
Then, it comes that
\begin{align}
cov \{ \Gamma _s^\alpha (k)\} = 2diag\{0,P_{\Gamma,\alpha}\}\label{25}
\end{align}
and
\begin{align}
&{cov} \{ {E_s}(k),\Gamma _s^\alpha (k)\}  + {cov} \{ \Gamma _s^\alpha (k),{E_s}(k)\} \nonumber\\
 &=  - 2\left[ {\begin{array}{*{20}{c}}
0&0\\
0&{{H_{w,\alpha  - 1}}{Q_{\alpha  - 1}}H_{w,\alpha  - 1}^T + {H_{v,\alpha  - 1}}{R_{\alpha  - 1}}H_{v,\alpha  - 1}^T}
\end{array}} \right]\nonumber\\
 &- \left[ {\begin{array}{*{20}{c}}
0&{{H_{w,s - \alpha }}{P_{wx,\alpha }}H_{0,\alpha  - 1}^T}\\
*&{{H_{{w_{21}},\alpha  - 1}}{P_{wx,\alpha }}H_{0,\alpha  - 1}^T + {H_{0,\alpha  - 1}}P_{wx,\alpha }^TH_{{w_{21}},\alpha  - 1}^T}
\end{array}} \right]\nonumber\\
 &- \left[ {\begin{array}{*{20}{c}}
0&{{H_{v,s - \alpha }}{P_{vx,\alpha }}H_{0,\alpha  - 1}^T}\\
*&0
\end{array}} \right],\label{26}
\end{align}
where
\begin{align}
{P_x} =~ &{cov} \{x(k)\},\nonumber\\
{P_{wx,\alpha }} =~& {cov} \{{W_{s - \alpha }}(k - \alpha ),x(k - \alpha  + 1)\},\nonumber\\
{P_{vx,\alpha }} =~& {cov} \{{V_{s - \alpha }}(k - \alpha ),x(k - \alpha  + 1)\},\nonumber\\
{P_{\Gamma,\alpha}}=~&{H_{0,\alpha  - 1}}{P_x}H_{0,\alpha  - 1}^T + {H_{w,\alpha  - 1}}{Q_{\alpha  - 1}}H_{w,\alpha  - 1}^T \nonumber\\
&+ {H_{v,\alpha  - 1}}{R_{\alpha  - 1}}H_{v,\alpha  - 1}^T,\nonumber
\end{align}
and $H_{w_{21},\alpha -1}$ is the lower left block of $H_{w,s}$ with row range $p(s+1-\alpha):p(s+1)$ and column range $1:q(s+1-\alpha)$.

Taking \eqref{25} and \eqref{26} into $\Delta_{\alpha}$, it comes that
\begin{align}
{\Delta _\alpha } = {Z_s}{P_{{\Delta _\alpha }}}Z_s^T,\label{27}
\end{align}
where
\begin{align}
&{P_{{\Delta _\alpha }}} =\left[ {\begin{array}{*{20}{c}}
0&0\\
0&{2{H_{0,\alpha  - 1}}{P_x}H_{0,\alpha  - 1}^T}
\end{array}} \right]\nonumber\\
 &- \left[ {\begin{array}{*{20}{c}}
0&{{H_{w,s - \alpha }}{P_{wx,\alpha }}H_{0,\alpha  - 1}^T}\\
*&{{H_{{w_{21}},\alpha  - 1}}{P_{wx,\alpha }}H_{0,\alpha  - 1}^T + {H_{0,\alpha  - 1}}P_{wx,\alpha }^TH_{{w_{21}},\alpha  - 1}^T}
\end{array}} \right]\nonumber\\
 &- \left[ {\begin{array}{*{20}{c}}
0&{{H_{v,s - \alpha }}{P_{vx,\alpha }}H_{0,\alpha  - 1}^T}\\
*&0
\end{array}} \right]\nonumber
\end{align}

\subsubsection{Residual Covariance Change with System Output Replay}
Consider the system output replay scenario as for \eqref{23}. Since
\begin{displaymath}
{cov} \{ {E_s}(k),{Y_{\alpha  - 1}}({\tau _k})\}  = 0,
\end{displaymath}
\begin{displaymath}
{cov} \{ {E_s}(k),{U_{\alpha  - 1}}({\tau _k})\}  = 0,
\end{displaymath}
\begin{displaymath}
{cov} \{ \hat x(k),\hat x({\tau _k})\}  = 0,
\end{displaymath}
it comes that
\begin{align}
&{cov} \left\{ {\Gamma _{s,u}^\alpha (k)} \right\}\nonumber\\
 &= 2\left[ {\begin{array}{*{20}{c}}
0&0\\
0&{{H_{u,\alpha  - 1}}{H_{c,\alpha  - 1}}{P_{\hat x}}H_{c,\alpha  - 1}^TH_{u,\alpha  - 1}^T}
\end{array}} \right],\label{28}
\end{align}
\begin{align}
&{cov} \left\{ {\Gamma _s^\alpha (k),\Gamma _{s,u}^\alpha (k)} \right\} + {cov} \left\{ {\Gamma _{s,u}^\alpha (k),\Gamma _s^\alpha (k)} \right\}\nonumber\\
 &= 2\left[ {\begin{array}{*{20}{c}}
0&0\\
0&{{H_{0,\alpha  - 1}}{P_{x,\hat x}}H_{c,\alpha  - 1}^TH_{u,\alpha  - 1}^T}
\end{array}} \right]\nonumber\\
&~~~ + 2\left[ {\begin{array}{*{20}{c}}
0&0\\
0&{{H_{u,\alpha  - 1}}{H_{c,\alpha  - 1}}{P_{\hat x,x}}H_{0,\alpha  - 1}^T}
\end{array}} \right],\label{29}
\end{align}
\begin{align}
&{cov} \left\{ {{E_s}(k),\Gamma _{s,u}^\alpha (k)} \right\} + {cov} \left\{ {\Gamma _{s,u}^\alpha (k),{E_s}(k)} \right\}\nonumber\\
 &=  - \left[ {\begin{array}{*{20}{c}}
0&{{H_{w,s - \alpha }}{P_{w\hat x,\alpha }}H_{c,\alpha  - 1}^TH_{u,\alpha  - 1}^T}\\
*&{\left\{ \begin{array}{l}
{H_{{w_{21}},\alpha  - 1}}{P_{w\hat x,\alpha }}H_{c,\alpha  - 1}^TH_{u,\alpha  - 1}^T\\
 + {H_{u,\alpha  - 1}}{H_{c,\alpha  - 1}}P_{w\hat x,\alpha }^TH_{{w_{21}},\alpha  - 1}^T
\end{array} \right\}}
\end{array}} \right]\nonumber\\
& ~~~~- \left[ {\begin{array}{*{20}{c}}
0&{{H_{v,s - \alpha }}{P_{v\hat x,\alpha }}H_{c,\alpha  - 1}^TH_{u,\alpha  - 1}^T}\\
*&0
\end{array}} \right],\label{30}
\end{align}
where
\begin{align}
{P_{\hat x}} =&~ {cov} \{\hat x(k)\},\nonumber\\
{P_{w\hat x,\alpha }} =&~ {cov} \{{W_{s - \alpha }}(k - \alpha ),\hat x(k - \alpha  + 1)\},\nonumber\\
{P_{v\hat x,\alpha }} =&~ {cov} \{{V_{s - \alpha }}(k - \alpha ),\hat x(k - \alpha  + 1)\}.\nonumber
\end{align}

Combining \eqref{28}, \eqref{29}, \eqref{30}, and \eqref{27} with \eqref{24} yields
\begin{align}
{\Delta _\alpha } = {Z_s}{P_{{\Delta _\alpha }}}Z_s^T + {Z_s}{P_{u,{\Delta _\alpha }}}Z_s^T,\label{31}
\end{align}
where
\begin{align}
&{P_{u,{\Delta _\alpha }}} = 2\left[ {\begin{array}{*{20}{c}}
0&0\\
0&{{H_{u,\alpha  - 1}}{H_{c,\alpha  - 1}}{P_{\hat x}}H_{c,\alpha  - 1}^TH_{u,\alpha  - 1}^T}
\end{array}} \right]\nonumber\\
& + 2\left[ {\begin{array}{*{20}{c}}
0&0\\
0&{{H_{0,\alpha  - 1}}{P_{x,\hat x}}H_{c,\alpha  - 1}^TH_{u,\alpha  - 1}^T}
\end{array}} \right]\nonumber\\
 &+ 2\left[ {\begin{array}{*{20}{c}}
0&0\\
0&{{H_{u,\alpha  - 1}}{H_{c,\alpha  - 1}}{P_{\hat x,x}}H_{0,\alpha  - 1}^T}
\end{array}} \right]\nonumber\\
 &- \left[ {\begin{array}{*{20}{c}}
0&{{H_{w,s - \alpha }}{P_{w\hat x,\alpha }}H_{c,\alpha  - 1}^TH_{u,\alpha  - 1}^T}\\
*&{\left\{ \begin{array}{l}
{H_{{w_{21}},\alpha  - 1}}{P_{w\hat x,\alpha }}H_{c,\alpha  - 1}^TH_{u,\alpha  - 1}^T\\
 + {H_{u,\alpha  - 1}}{H_{c,\alpha  - 1}}P_{w\hat x,\alpha }^TH_{{w_{21}},\alpha  - 1}^T
\end{array} \right\}}
\end{array}} \right]\nonumber\\
& - \left[ {\begin{array}{*{20}{c}}
0&{{H_{v,s - \alpha }}{P_{v\hat x,\alpha }}H_{c,\alpha  - 1}^TH_{u,\alpha  - 1}^T}\\
*&0
\end{array}} \right].\label{32}
\end{align}

\subsubsection{Interpretation and Discussion of the Residual Covariance Change}
It is of interest to study $\Delta_{\alpha}$ in \eqref{27} and \eqref{31} to reveal the possibility of stealthy attack implementation and successful attack detection. Both system state and dynamics are involved in $\Delta_{\alpha}$, and it is concluded that better closed-loop control performance, e.g., smaller $\|P_x\|$ and $\|P_{\hat x}\|$, implies lower attack detection rate and higher stealthiness for attack implementation. This is due to the fast convergence effect of the closed-loop system, which usually implies a weak cross-correlation between system input and output data series within a short time lag.

Regarding the occurrence of replay attack, it is known that $\Delta_{\alpha} \ne 0$ when $0<\alpha \le s$ based on Proposition 2. How about the residual change due to the end or disappearance of replay attack? It can be derived that the non-zero covariance change of residual holds $s$ steps starting from the stop of replay attack, according to the finite horizon property of the parity space scheme as well as the proof of Proposition 2. Thus, both the occurrence and disappearance of the replay attack show a finite time tailing effect (or covariance change) on the parity space based residual. Based on this property of $\Delta_{\alpha}$, the distinction between fault and replay attack can be achieved directly, where the permanent fault causes no tailing effect on the residual and the intermittent fault causes no tailing effect holding exactly the same length and type as that of the replay attack.

Moreover, when $\alpha$ is relatively small, it is known from \eqref{27} and \eqref{32} that there are few nonzero elements in the weighting matrix ${P_{{\Delta _\alpha }}}$ and ${P_{u,{\Delta _\alpha }}}$ for the quadratic forms with respect to $Z_s$, which implies a relative small $tr\left( {{\Delta _\alpha }} \right)$. However, when $\alpha$ approaches $s+1$, we can conclude that $tr\left( {{\Delta _\alpha }} \right)$ will become relative small again due to the fact of ${Z_s}{H_{0,s}} = 0$ and both ${H_{u,\alpha  - 1}}$ and ${H_{0,\alpha  - 1}}$ contain part of ${H_{0,s}}$. Thus, the replay attack effect will increase to a peak and then unsymmetrically decrease to the nominal value, which also facilitate the distinction between fault and replay attack. The detailed analysis of distinction between system fault and replay attack for anomaly detection is given in the next Section.

Since a covariance change is caused, replay attack can be viewed as a special multiplicative fault to the system. However, the change of the covariance is indefinite since the weighting of $\Delta_{\alpha}$ in both \eqref{27} and \eqref{31}, i.e., ${P_{{\Delta _\alpha }}}$ and ${P_{{\Delta _\alpha }}} + {P_{u,{\Delta _\alpha }}}$, are indefinite \cite{24_zhang2005,25_ferrari2019,26_horn2012}. Note that the covariance change is contributed by $P_x$, $P_{\hat x}$, $P_{x,\hat x}$, $P_{wx,\alpha}$, $P_{vx,\alpha}$, $P_{w\hat x,\alpha}$, and $P_{v\hat x,\alpha}$. Since the closed-loop system is stable, the norms of $P_{x,\hat x}$, $P_{wx,\alpha}$, $P_{vx,\alpha}$, $P_{w\hat x,\alpha}$, and $P_{v\hat x,\alpha}$ are usually much smaller than those of $P_x$ and $P_{\hat x}$.

\textbf{Remark 4}. \emph{The existence of $\Delta_{\alpha}$ has been explained above; however, it is of interest to emphasize the key reason for the non-zero $\Delta_{\alpha}$. Since a finite horizon detection scheme is used, we retrospect the control performance change within finite steps just after the occurrence of replay attack, which is an in-time detection and avoids the change (or ``ripples") vanishment due to infinite horizon scheme, e.g., consider the performance change for $k \to \infty$.}

\subsection{Replay Attack Detection and Detectability with the Parity Space based Residual}
Based on the analysis in Section III.B, it is known that a covariance change of the residual is triggered by the replay attack. Thus, different test statistics can be adopted for replay attack detection. In the following, the most used $\chi ^2$ test and likelihood ratio (LR) based replay attack detection scheme and detectability analysis will be presented.

\subsubsection{$\chi ^2$ Test Based Replay Attack Detection and Detectability}
Based on \eqref{4}, define the following test statistic
\begin{align}
{J_{s,{\chi ^2}}}(k) = r_s^T(k)\Theta _s^{ - 1}{r_s}(k),\label{33}
\end{align}
then it is known that ${J_{s,{\chi ^2}}}(k) \sim {\chi ^2}(l)$. Thus, based on a given false alarm rate $\gamma$, the detection threshold ${J_{th,}}_{{\chi ^2}}$ can be determined by using the available $\chi ^2$ data table \cite{27_ding2014}. With ${J_{s,{\chi ^2}}}(k)$ and ${J_{th,}}_{{\chi ^2}}$, the following decision logic is established:
\begin{equation}\label{34}
\left\{ \begin{array}{l}
{J_{s,{\chi ^2}}}(k) \le {J_{th,}}_{{\chi ^2}},~attack~free,\\
{J_{s,{\chi ^2}}}(k) > {J_{th,}}_{{\chi ^2}},~attack~on.
\end{array} \right.
\end{equation}
In the presence of replay attack, $r_s(k)$ is replaced by $r_s^{\alpha}(k)$ in practice and the test statistic turns to be
\begin{align}
J_{s,{\chi ^2}}^\alpha (k) = {\left( {r_s^\alpha (k)} \right)^T}\Theta _s^{ - 1}r_s^\alpha (k).\label{35}
\end{align}
To check the detectability of the replay attack based on the pair $\left\{ {J_{s,{\chi ^2}}^\alpha (k),{J_{th,}}_{{\chi ^2}}} \right\}$, the following Theorem is given.

\textbf{Theorem 1}. \emph{With the given threshold ${J_{th,}}_{{\chi ^2}}$, test statistic ${J_{s,{\chi ^2}}}(k)$ in \eqref{33}, and decision logic in \eqref{34},
\begin{enumerate}
  \item the replay attack detection rate is upper bounded by ${\bar \mu _{{\chi ^2}}}$;
  \item  if $l + tr\left( {{\Delta _\alpha }\Theta _s^{ - 1}} \right) - {J_{th,{\chi ^2}}} \ge 0$, the replay attack detection rate is lower bounded by ${\underline{\mu}_{{\chi ^2}}}$;
\end{enumerate}
where
\begin{align}
{\bar \mu _{{\chi ^2}}} = \min \left( {\frac{{l + tr\left( {{\Delta _\alpha }\Theta _s^{ - 1}} \right)}}{{{J_{th,}}_{{\chi ^2}}}},1} \right),\label{36}
\end{align}
\begin{align}
\underline{\mu}_{\chi ^2}= \frac{{\bar \kappa _\chi ^2}}{{1 + \bar \kappa _\chi ^2}},\label{37}
\end{align}
\begin{displaymath}
{\bar \kappa _\chi } = \frac{{l + tr\left( {{\Delta _\alpha }\Theta _s^{ - 1}} \right) - {J_{th,{\chi ^2}}}}}{{2tr(\Theta _s^{ - 1}\Theta _s^\alpha \Theta _s^{ - 1}\Theta _s^\alpha )}}.
\end{displaymath}}

\begin{proof}
Based on \eqref{18}, the mean of $J_{s,{\chi ^2}}^\alpha (k)$ is derived as
\begin{align}
{\mathcal{E}}\{J_{s,{\chi ^2}}^\alpha (k) \}=~& {\mathcal{E}}\big\{ {{{\left( {r_s^\alpha (k)} \right)}^T}\Theta _s^{ - 1}r_s^\alpha (k)} \big\}\nonumber\\
 =~& l + tr\left( {{\Delta _\alpha }\Theta _s^{ - 1}} \right).\label{38}
\end{align}
Since $J_{s,{\chi ^2}}^\alpha (k) \ge 0$, based on the Markov inequality and ${\mathcal{E}}\{J_{s,{\chi ^2}}^\alpha (k)\}$ given in \eqref{38}, it comes that
\begin{align}
P\left( {J_{s,{\chi ^2}}^\alpha (k) \ge {J_{th,}}_{{\chi ^2}}} \right) \le {\bar \mu _{{\chi ^2}}},\label{39}
\end{align}
where ${\bar \mu _{{\chi ^2}}}$ is given in \eqref{36}.

Based on Lemma 2 of \cite{28_fu2022}, a conditional lower bound of the replay attack detection probability $\underline{\mu}_{\chi ^2}$ is given by
\begin{align}
P\left(J_{s,\chi ^2}^\alpha (k) \ge J_{th,\chi ^2}\right)\ge\underline{\mu}_{\chi ^2},\label{40}
\end{align}
if
\begin{displaymath}
l + tr\left( {{\Delta _\alpha }\Theta _s^{ - 1}} \right) - {J_{th,{\chi ^2}}} \ge 2\kappa tr(\Theta _s^{ - 1}\Theta _s^\alpha \Theta _s^{ - 1}\Theta _s^\alpha ),
\end{displaymath}
where $\kappa = \sqrt{\frac{\underline{\mu}_{\chi ^2}}{1-\underline{\mu}_{\chi ^2}}}$. Since ${cov}\{J_{s,{\chi ^2}}^\alpha (k) \} > 0$ and $l + tr\left( {{\Delta _\alpha }\Theta _s^{ - 1}} \right) - {J_{th,{\chi ^2}}} \ge 0$, we have $\underline{\mu}_{\chi ^2}$ as given in \eqref{37}. Finally, Theorem 1 is proved.
\end{proof}

Note that $\bar \mu _{\chi ^2}$ is the upper bound of the replay attack detection rate when parity space formulation and $\chi ^2$ test are used. Specifically, if $l + tr\left( {{\Delta _\alpha }\Theta _s^{ - 1}} \right) > {J_{th,}}_{{\chi ^2}}$, ${\bar \mu _{{\chi ^2}}} = 1$ and $\underline{\mu}_{\chi ^2}>0$, which supports the detectability result in Theorem 1; if $l + tr\left( {{\Delta _\alpha }\Theta _s^{ - 1}} \right) < {J_{th,}}_{{\chi ^2}}$, ${\bar \mu _{{\chi ^2}}} < 1$ and the condition for \eqref{37} failed, which also means no guarantee for the lower bound of the detection rate. Thus, when $tr\left( {{\Delta _\alpha }\Theta _s^{ - 1}} \right)$ is large enough, a lower bound of the detection rate can be ensured; when $tr\left( {{\Delta _\alpha }\Theta _s^{ - 1}} \right)$ is relative small, the upper bound of the detection rate can be known from \eqref{39}. Since $J_{s,{\chi ^2}}^\alpha (k)$ does not belong to $\chi ^2(l)$ when $0<\alpha \le s$ due to the presence of replay attack (or ${\Delta _\alpha }$), it is hard to get the exact probability of detection performance by referring to the probability distribution table directly. Thus, the results in \eqref{36} and \eqref{37} together present the probability range of detection performance. Due to the lack of full distribution information of $J_{s,{\chi ^2}}^\alpha (k)$, the range in Theorem 1 may be conservative.

\textbf{Remark 5}. \emph{For diagnosability analysis, specific feature of test statistic is concerned. If the mean of the test statistic is considered, based on \eqref{38}, the detectability condition of replay attack is
\begin{align}
tr\left( {{\Delta _\alpha }\Theta _s^{ - 1}} \right) > 0, \label{41}
\end{align}
as the test statistic in \eqref{33} is in the form of $T^2$. However, the condition in \eqref{41} does not always imply a successful replay attack in practice, as it indicates only the feasibility of detection with an potential increase of the test statistic. This gap, between detectability condition and detection performance, has been discussed in \cite{28_fu2022}. In this case, we present the quantitative probability range of detection for replay attack detection performance evaluation in practice. Also, if the covariance of the test statistic is considered, the corresponding detectability can be derived, and the Riemannian manifold as well as symmetric positive definite matrix theory can be adopted to measure the detectability of replay attack \cite{29_ding2021}.}

\subsubsection{LR Based Replay Attack Detection and Detectability}
It is known that the $\chi ^2$ test is efficient for mean change or additive anomaly detection \cite{29_ding2021}. To ensure a successful detection of replay attack, the multiplicative anomaly effect on the residual, i.e., $\left\| {\Gamma _s^\alpha (k)} \right\|$ as well as $\left\| {\Gamma _{s,u}^\alpha (k)} \right\|$, should be significantly large compared to $\left\| {{r_s}(k)} \right\|$ when $\chi ^2$ test is used; however, this magnitude requirement of anomaly effect may not hold. Considering this case, GLR test is adopted, and the performance of LR based replay attack detection is further analyzed.

Let $L(\varepsilon |{r_s}(k))$ be the likelihood function of the probability density function parameter set $\varepsilon$ given by $r_s(k)$. The likelihood ratio for replay attack detection is
\begin{align}
h({r_s}(k)) = \frac{{L({\varepsilon _a}|{r_s}(k))}}{{L({\varepsilon _0}|{r_s}(k))}},\label{42}
\end{align}
where $\varepsilon _0$ and $\varepsilon _a$ are the parameter sets without and with replay attack, respectively.

For replay attack detection, it is clear that
\begin{equation}\label{43}
\varepsilon  = \left\{ \begin{array}{l}
{\Theta _s},~without~attack,\\
\Theta _s^\alpha ,~with~attack.
\end{array} \right.
\end{equation}
For convenience, $log$-likelihood ratio is adopted, then
\begin{align}
h({r_s}(k)) = \ln \frac{{L(\Theta _s^\alpha |{r_s}(k))}}{{L({\Theta _s}|{r_s}(k))}}.\label{44}
\end{align}
Note that the replay attack occurrence time $T_0$ as well as the parameter set $\varepsilon_a$ is unknown for real-time detection. Thus, the GLR,
\begin{align}
\hat h({r_s}(k)) = \ln \frac{{L(\hat \Theta _s^\alpha |{r_s}(k))}}{{L({\Theta _s}|{r_s}(k))}}, \label{45}
\end{align}
is adopted, where $\hat \Theta _s^\alpha $ is the maximum likelihood estimation of $\Theta _s^\alpha $. Since $w(k)$ and $v(k)$ are Gaussian white noises, it is known that $r_s(k)$ is Gaussian when attack free. The probability density function of $r_s(k)$ is
\begin{align}
{f_\varepsilon }({r_s}(k)) = \frac{1}{{\sqrt {{{(2\pi )}^l}\det \varepsilon } }}{e^{ - \frac{1}{2}r_s^T(k){\varepsilon ^{ - 1}}{r_s}(k)}}.\label{46}
\end{align}
With $n_r$ observations of $r_s(k)$, the test statistic for implementation based on \eqref{45} and \eqref{46} is given by
\begin{align}
{J_{{n_r},LR}}(k) =~& \frac{1}{2}{n_r}\ln \frac{{\det {\Theta _s}}}{{\det \hat \Theta _s^\alpha }}\nonumber\\
 &+ \frac{1}{2}\sum\limits_{i = 0}^{{n_r} - 1} {} \big\{ r_s^T(k - i)\Theta _s^{ - 1}{r_s}(k - i)\nonumber\\
 &- r_s^T(k - i){(\hat \Theta _s^\alpha )^{ - 1}}{r_s}(k - i)\big\},\label{47}
\end{align}
where $\hat \Theta _s^\alpha  = \frac{1}{{{n_r}}}\sum\limits_{i = 0}^{{n_r} - 1} {\left( {{r_s}(k - i)r_s^T(k - i)} \right)}$.

The threshold $J_{th,LR}$ design for the GLR based detection is not a routine task, due to the lack of analytical solution for the probability computation. The numerical solutions have been well studied and adopted for such a situation. For simplicity, please refer to \cite{29_ding2021} (Section 15.1).

For detectability performance analysis, as both ${\Theta _s}$ and $\Theta _s^\alpha $ are known, the test statistic $h({r_s}(k))$ is considered, and the LR based replay attack detectability is summarized in the following Theorem.

\textbf{Theorem 2}. \emph{With the given threshold $J_{th,LR}$ and test statistic in \eqref{44},
\begin{enumerate}
  \item the replay attack detection rate is upper bounded by $\bar \mu _{LR}$;
  \item if $\frac{1}{2}\left( {\ln \frac{{\det {\Theta _s}}}{{\det \Theta _s^\alpha }} + tr\left( {{\Delta _\alpha }\Theta _s^{ - 1}} \right)} \right) - {J_{th,LR}} \ge 0$, the replay attack detection rate is lower bounded by $\underline \mu _{LR}$,
\end{enumerate}
where
\begin{align}
{\bar \mu _{LR}} = \min \left( {\frac{{\ln \frac{{\det {\Theta _s}}}{{\det \Theta _s^\alpha }} + tr\left( {{\Delta _\alpha }\Theta _s^{ - 1}} \right)}}{{2{J_{th,GLR}}}},1} \right),\label{48}
\end{align}
\begin{align}
\underline \mu_{LR}=\frac{{\bar \kappa _{LR}^2}}{{1 + \bar \kappa _{LR}^2}},\label{49}
\end{align}
and
\begin{displaymath}
{\bar \kappa _{LR}} = \frac{{\ln \frac{{\det {\Theta _s}}}{{\det \Theta _s^\alpha }} + tr\left( {{\Delta _\alpha }\Theta _s^{ - 1}} \right) - 2{J_{th,LR}}}}{{4tr\{ {{(\Theta _s^{ - 1}\Theta _s^\alpha )}^2}\}  + 4l - 8tr\{ \Theta _s^{ - 1}\Theta _s^\alpha \} }}.
\end{displaymath}}

\begin{proof}
After the occurrence of the replay attack, it is known that
\begin{align}
\mathcal{E}\{ h({r_s}(k))\}  =~& D({f_{\Theta _s^\alpha }}({r_s}(k)),{f_{{\Theta _s}}}({r_s}(k)))\nonumber\\
 =~& \frac{1}{2}\left( {\ln \frac{{\det {\Theta _s}}}{{\det \Theta _s^\alpha }} + tr\left( {{\Delta _\alpha }\Theta _s^{ - 1}} \right)} \right),\label{50}
\end{align}
where $D({f_{\Theta _s^\alpha }}({r_s}(k)),{f_{{\Theta _s}}}({r_s}(k)))$ is the Kullback-Leibler divergence of the distributions of residual with and without replay attack \cite{29_ding2021}.

Similar to the proof for Theorem 1, $\bar \mu_{LR}$ is obtained based on $\mathcal{E}\{h(r_s(k))\}$ in \eqref{50} and the given threshold $J_{th,LR}$.

Based on Lemma 2 of \cite{28_fu2022}, a conditional lower bound of the replay attack detection probability $\underline \mu_{LR}$ is given by
\begin{align}
P\left( {h({r_s}(k)) \ge {J_{th,LR}}} \right) \ge \underline \mu_{LR},\label{51}
\end{align}
if $D({f_{\Theta _s^\alpha }}({r_s}(k)),{f_{{\Theta _s}}}({r_s}(k))) - {J_{th,LR}} \ge {\kappa _{LR}}{cov} (h({r_s}(k)))$, where $\kappa_{LR}= \sqrt{\frac{\underline \mu_{LR}}{1-\underline \mu_{LR}}}$. With \eqref{50} and $cov(h(r_s(k)))$ given in Appendix, $\underline \mu_{LR}$ is determined as given in \eqref{49}. Finally, Theorem 2 is proved.
\end{proof}

Note that the result in Theorem 2 is an ideal case for the GLR based detection performance since $\Theta _s^\alpha $ is used for detectability analysis.

\textbf{Remark 6}. \emph{Similar to the detectability analysis in Remark 5, the replay attack detectability condition is
\begin{align}
\ln \frac{{\det {\Theta _s}}}{{\det \Theta _s^\alpha }} + tr\left( {{\Delta _\alpha }\Theta _s^{ - 1}} \right) > 0.\label{52}
\end{align}
Kullback-Leibler divergence is adopted for detectability quantification. It is obviously that \eqref{52} is identically true \cite{29_ding2021}, because the LR based detection means to compare the residual distributions with and without replay attack. Nevertheless, one has to enlarge the left side of \eqref{52} to improve the detection performance, which will be studied in the next Section.}

\textbf{Remark 7}. \emph{To check the detectability condition and the detection performance bound for both $\chi ^2$ and $LR$ based replay attack detection, the key is to get $\Delta_{\alpha}$. Besides, $\Delta_{\alpha}$ can be used to measure the intrinsic security of the considered control system against replay attack for both the defender and attacker. Instead of theoretical calculation (introduced in Section IV.A.3), one can``learn" $cov \{ {Z_s}\Gamma _s^\alpha (k)\} $ and $\Delta_{\alpha}$ based on the attack input and output data sets from the detector side. As ${\Theta _s}$ is known (or can be learnt), $\hat {\Delta}_{\alpha}$, the estimation of ${\Delta}_{\alpha}$, is given by
\begin{displaymath}
{\hat \Delta _\alpha } = \hat \Theta _s^\alpha  - {\Theta _s},
\end{displaymath}
where $\hat \Theta _s^\alpha $ is the estimation of $cov \left\{ {r_s^\alpha (k)} \right\}$ and can be obtained by referring to Algorithm 1, which will be given in Section IV.A.3.}

\subsection{Security of Parity Space Based Replay Attack Detection Scheme}
Based on the detection performance analysis, one can find that the replay attack is not absolutely stealthy when the parity space based detection is employed. Thus, it is important for the attacker to estimate the stealthiness level of the replay attack in advance to avoid potential alarm. Specifically, the detection performance for the detector is inversely proportional to the stealthiness level for the attacker. Based on this fact, the attacker can identify the vulnerable control system and decide the stealthy starting time period for attack.

The stealthiness evaluation of control system for the attacker can be done in stage-$1$ of the replay attack and both system input and output data is required. Again, the key is to get $\Theta _s^\alpha $ or $Z_s$. In stage-$1$ of the replay attack, data-driven parity vector identification is feasible based on system input and output data \cite{27_ding2014}. With the estimation of the parity vector and input and output data, $\Theta _s^\alpha $ and ${\Theta _s}$ can be estimated with and without off-line data replay, respectively. Based on the estimation of $\Theta _s^\alpha $ and ${\Theta _s}$, the replay attack detection/stealthiness performance can be known based the results in Theorems 1 and 2.

Even so, due to the lack of dimension information of $Z_s$, the data window length $s$, and the detection threshold, the stealthiness level cannot be well estimated. From the perspective of cyber security, the window length $s$ and parity matrix $Z_s$ \emph{encrypt} the detector. To keep the security of the parity space based replay attack detection scheme, multiple residuals are suggested, where each residual corresponds to different $Z_s$, $s$, and detection threshold. Besides, since the essential output data replay induced data switching is inevitable for the replay attack implementation, the replay attack effect on residual is a kind of intrinsic property of the control system, determined together by the detector and the control system dynamics. Thus, the security level against replay attack for control systems is also partially determined by the dynamics.

Based on the analysis in Section III.C, the direction for improving the detection performance/decreasing the stealthiness level of replay attack is motivated: to enlarge the difference between ${\Theta _s}$ and $\Theta _s^\alpha $ for detection, which is introduced in Section IV.
\subsection{Fault and Replay Attack Distinction}
To distinguish system fault and replay attack, the comparison between system fault-driven residual and replay-attack-driven residual is given.

Assume system \eqref{1} to be faulty:
\begin{eqnarray}\nonumber
\left\{ \begin{array}{l}
x(k + 1) = Ax(k) + {B_u}u(k) + {B_w}w(k)+ {B_f}f(k),\\
y(k) = Cx(k) + v(k)+{D_f}f(k),
\end{array} \right.
\end{eqnarray}
where $f(k) \in \mathcal{R}^{n_f}$ is system fault and $B_f$ and $D_f$ are fault distribution matrices with appropriate dimensions. With different $B_f$ and $D_f$, $f(k)$ can be used to model actuator fault, process fault, and sensor fault \cite{23_ding2008}.

Under the existence of fault $f(k)$, the parity space based residual turns to be
\begin{align}
{r_s}(k) = {Z_s}{E_s}(k)+Z_sH_{f,s}F_s(k), \label{d1}
\end{align}
where $H_{f,s}$ and $F_s(k)$ can be obtained based on $H_{w,s}$ and $Y_s(k)$, respectively.

Generally, in spite of the fault location, $f(k)$ can be deterministic or statistic and time-varying or time-invariant. But, the concerned effect of $F_s(k)$ on the residual is its mean change and/or covariance change. Note that $\mathcal{E}\{E_s(k)\}=0$ and $cov\{Z_sE_s(k)\}=\Theta_s$. Thus, the residual mean and covariance changes induced by the nonzero $F_s(k)$ implies that $\mathcal{E}\{r_s(k)\} \ne 0$ and $cov\{r_s(k)\} \ne \Theta_s$, with $r_s(k)$ given in \eqref{d1}.

One dimensional test statistic is adopted, i.e., the $\chi ^2$ test statistic in \eqref{33} or the GLR test statistic in \eqref{45}. The change from $\mathcal{E}\{E_s(k)\}=0$ to $\mathcal{E}\{r_s(k)\} \ne 0$ will enlarge the test statistic in \eqref{33} and in \eqref{45}, compared to the fault-free case; the change from $cov\{Z_sE_s(k)\}=\Theta_s$ to $cov\{r_s(k)\} \ne \Theta_s$ will enlarge the test statistic in \eqref{45} and decrease or enlarge the test statistic in \eqref{33}, compared to the fault-free case. Thus, from the detection perspective, it is known that system fault, just like replay attack, causes a change on the adopted test statistic based on parity space method, despite the original change form (on the residual mean or covariance).

Note that the duration of the statistical property change of the residual caused by system fault is determined by system fault type; however, the duration of covariance change of residual due to replay attack is determined by the order of parity relation, when fault and replay attack last longer than $s$ steps. Moreover, both the occurrence and disappearance of replay attack cause residual covariance change within $s$ steps. Also, the replay attack driven residual covariance change will make an inverted v-shape change of the adopted test statistic with respect to the order of parity relation $s$, if the replay attack can be detected. Thus, by using the information of the residual response pattern and the order of parity relation, the theoretical features of fault and replay attack on $\chi ^2$ and GLR test statistics can be distinguished. To enhance the distinction performance between system fault and replay attack, observer-based detector is recommended to run in parallel with the parity space based detector, since the classical observer-based or Kalman-filter-based scheme is not sensitive to replay attack as reported in \cite{3_dibaji2019,5_sanchez2019,6_mo2009,18_ding2022}. In addition to the observer-based detection method, as mentioned in Section III.D, multiple residual generations are suggested with different choice of $s$ to deal with the sparse fault and replay attack distinction.

\textbf{Remark 8}. \emph{The rare case, one faulty point or outlier lies in the detection window, can cause a residual change lasting for $s$ steps, which is similar to the case of replay attack. However, the one outlier case can be catched more easily by the observer-based detection scheme than the parity space based detection scheme if it is detectable, and the residual change response pattern is different from the replay attack case except the response duration. Thus, the sparse outlier can be distinguished from the replay attack as well.}

\section{Performance Enhancement for Parity Space Method Based Replay Attack Detection}
In this Section, both passive and active design methods are proposed to improve the replay attack detection performance, where the passive design aims to optimize the detection performance without any plant-side modification and the active design means a targeted plant-side modification to facilitate the detection of replay attack.

\subsection{Passive Design}
Based on the characterized residual change under replay attack, the optimization of the parity matrix $Z_s$ is proposed to improve the detection performance.
\subsubsection{Optimal Parity Matrix}
The parity matrix $Z_s$ can be parameterized as
\begin{align}
{Z_s} = {M_s}{N_s},\label{53}
\end{align}
where $N_s H_{0,s}=0$, $N_s \ne0$, and $M_s \in \mathcal{R}^{{l \times {n_z}}}$ is the nonzero weighting matrix to be designed.

As motivated by Theorem 2, the key to improve the detection performance is to make sure that $\Theta _s^\alpha $ or ${\Delta _\alpha }$ is relatively different from $\Theta_s$. Note that the weighting matrices for ${\Delta _\alpha }$, $P_{{\Delta _\alpha }}$ and/or $P_{u,{\Delta _\alpha }}$, are indefinite. There are various ``directions" to go to enlarge the difference between $\Theta _s^\alpha $ and ${\Theta _s}$. Under the circumstances, four general optimization problems will be addressed for replay attack detection regarding different indices.

\emph{Positive definite portion optimization}. Let
\begin{align}
P_{x,P}^\alpha  =& \left[ {\begin{array}{*{20}{c}}
0&0\\
*&{2{H_{0,\alpha  - 1}}{P_x}H_{0,\alpha  - 1}^T}
\end{array}} \right],\nonumber\\
P_{\hat x,P}^\alpha  =& \left[ {\begin{array}{*{20}{c}}
0&0\\
*&{{H_{u,\alpha  - 1}}{H_{c,\alpha  - 1}}{P_{\hat x}}H_{c,\alpha  - 1}^TH_{u,\alpha  - 1}^T}
\end{array}} \right],\nonumber
\end{align}
then it is known that ${Z_s}P_{x,P}^\alpha Z_s^T$ and ${Z_s}P_{\hat x,P}^\alpha Z_s^T$ are semi-positive definite portions in $\Delta_{\alpha}$. Thus, we can maximize these semi-positive definite portions compared to ${Z_s}{\Theta _s}Z_s^T$.

Let
\begin{displaymath}
{T_{{\Theta _s}}} = {N_s}\left( {{H_{w,s}}{Q_s}H_{w,s}^T + {H_{v,s}}{R_s}H_{v,s}^T} \right)N_s^T,\\
\end{displaymath}
\begin{displaymath}
{T_{{\Delta _\alpha },P}} = \left\{ \begin{array}{l}
\begin{aligned}
&{N_s}P_{x,P}^\alpha N_s^T,&with~input~replay,\\
&{N_s}\left( {P_{x,P}^\alpha  + P_{\hat x,P}^\alpha } \right)N_s^T,&without~input~replay,
\end{aligned}
\end{array} \right.
\end{displaymath}
then we have
\begin{equation}\label{54}
\left\{ \begin{array}{l}
{\Theta _s} = {M_s}{T_{{\Theta _s}}}M_s^T,\\
{\Delta _{\alpha ,P}} = {M_s}{T_{{\Delta _\alpha },P}}M_s^T.
\end{array} \right.
\end{equation}

Consider the following ratio type index
\begin{equation}\label{55}
{J_1} = \frac{{{{\left\| {{M_s}T_{{\Delta _\alpha },P}^{\frac{1}{2}}} \right\|}_i}}}{{{{\left\| {{M_s}T_{{\Theta _s}}^{\frac{1}{2}}} \right\|}_i}}},~i = 2,\infty.
\end{equation}
It is known that $T_{{\Theta _s}}^{\frac{1}{2}} \in {R^{{n_z} \times (q + p)(s + 1)}}$. Without loss of generality, we assume $T_{{\Theta _s}}^{\frac{1}{2}}$ a full row rank matrix. Therefore, doing SVD of $T_{{\Theta _s}}^{\frac{1}{2}}$ yields
\begin{equation}\label{56}
T_{{\Theta _s}}^{\frac{1}{2}} = U\left[ {\begin{array}{*{20}{c}}
\Lambda &0
\end{array}} \right]{V^T},
\end{equation}
where $UU^T=I$ and $VV^T=I$. The solution to the maximization problem of $J_1$ in \eqref{55} with respect to $M_s$ is summarized below.

\textbf{Theorem 3}. \emph{The solution
\begin{align}
{M_s} = {\Lambda ^{ - 1}}{U^T}\label{57}
\end{align}
solves the optimization problem $\mathop {\max }\limits_{{M_s}} {J_1}$ with maximum
\begin{align}
{\bar J_1} = {\left\| {{\Lambda ^{ - 1}}{U^T}T_{{\Delta _\alpha },P}^{\frac{1}{2}}} \right\|_i},~i = 2,\infty .\label{58}
\end{align}}

\begin{proof}
The proof for Theorem 3 is similar to the proof given in \cite{23_ding2008} (Chapter 7.4), and thus the details are omitted here.
\end{proof}

Note that the solution in \eqref{57} also solves
\begin{displaymath}
\mathop {\max }\limits_{{M_s}} \frac{{{\sigma _i}\left( {{M_s}T_{{\Delta _\alpha },P}^{\frac{1}{2}}} \right)}}{{{{\left\| {{M_s}T_{{\Theta _s}}^{\frac{1}{2}}} \right\|}_2}}}
\end{displaymath}
for all $i$. Since the weighting in \eqref{57} is independent of ${T_{{\Delta _\alpha },P}}$ or $\alpha$, it means to optimize the replay attack detection performance in terms of index $J_1$ at any steps. Thus, \eqref{57} is also the solution for the following optimization problem:
\begin{displaymath}
\mathop {\max }\limits_{{M_s}} \frac{{{{\left\| {{M_s}{{\left\{ {\sum\limits_{\alpha  = 1}^s {{T_{{\Delta _\alpha },P}}} } \right\}}^{\frac{1}{2}}}} \right\|}_i}}}{{{{\left\| {{M_s}T_{{\Theta _s}}^{\frac{1}{2}}} \right\|}_i}}},~i = 2,\infty ,
\end{displaymath}
where an interval accumulation index is adopted.

\emph{Ratio trace optimization}. For both the $\chi^2$ and GLR based replay attack detection, the term $tr\left( {\Theta _s^{ - 1}\Theta _s^\alpha } \right)$ (or $\Delta_{\alpha}$) is the key for maximizing the detectability of replay attack. Thus, we propose the following index
\begin{equation}\label{59}
{J_2} = \sum\limits_{\alpha {\rm{ = }}1}^s {tr\left( {\Theta _s^{ - 1}{\Delta _\alpha }} \right)} .
\end{equation}
The index $J_2$ is motivated by the fact that the replay attack effect is nonzero when $0<\alpha \le s$, and it requires an interval accumulation index optimization for detection.

Let
\begin{equation}\label{60}
{\Delta _{s,\Sigma }} = {M_s}{T_{\Delta ,\Sigma }}M_s^T,
\end{equation}
where
\begin{displaymath}
{T_{\Delta ,\Sigma }} = \sum\limits_{\alpha {\rm{ = }}1}^s {{T_{{\Delta _\alpha }}}} ,
\end{displaymath}
\begin{displaymath}
{T_{{\Delta _\alpha }}} = \left\{ \begin{array}{l}
\begin{aligned}
&{N_s}{P_{{\Delta _\alpha }}}N_s^T,&with~input~replay,\\
&{N_s}\left( {{P_{{\Delta _\alpha }}} + {P_{u,{\Delta _\alpha }}}} \right)N_s^T,&without~input~replay.
\end{aligned}
\end{array} \right.
\end{displaymath}
Taking \eqref{60} into \eqref{59} yields
\begin{align}
{J_2} =~& \sum\limits_{\alpha {\rm{ = }}1}^s {tr\left( {\Theta _s^{ - 1}{\Delta _\alpha }} \right)} \nonumber\\
 =~& tr\left( {\Theta _s^{ - 1}{M_s}{T_{\Delta ,\Sigma }}M_s^T} \right).\label{61}
\end{align}
The solution for optimizing $J_2$ in \eqref{61} is summarized in the following Theorem.

\textbf{Theorem 4}. \emph{Let $M_{s,i}^T$ and $\lambda_i$ be the $i$-th largest generalized eigenvector and eigenvalue for $\left( {{T_{\Delta ,\Sigma }},{T_{{\Theta _s}}}} \right)$, then
\begin{align}
{M_s} = \left[ {\begin{array}{*{20}{c}}
{{M_{s,1}}}\\
 \vdots \\
{{M_{s,l}}}
\end{array}} \right] \label{62}
\end{align}
solves the ratio trace problem $\mathop {\max }\limits_{{M_s}} {J_2}$ with maximum ${\bar J_2}{\rm{ = }}\sum\limits_{i = 1}^l {{\lambda _i}} $.}

\begin{proof}
Based on \cite{30_wang2007} and \cite{31_fukunaga2013}, it is known that
\begin{align}
\arg &\mathop {\max }\limits_{{M_s}} \frac{{\det \left( {{M_s}{T_{\Delta ,\Sigma }}M_s^T} \right)}}{{\det \left( {{M_s}{T_{{\Theta _s}}}M_s^T} \right)}} \nonumber\\
&= \arg \mathop {\max }\limits_{{M_s}} tr\left( {\Theta _s^{ - 1}{M_s}{T_{\Delta ,\Sigma }}M_s^T} \right)\label{63}
\end{align}
and the solution to the determinant ratio problem in \eqref{62} solves the ratio trace optimization problem. Based on the solution of $M_s$, we have ${T_{\Delta ,\Sigma }}M_{s,i}^T = {\lambda _i}{T_{{\Theta _s}}}M_{s,i}^T$, then
\begin{align}
{M_s}{T_{\Delta ,\Sigma }}M_s^T = {M_s}{T_{{\Theta _s}}}M_s^T\Xi ,\label{64}
\end{align}
where $\Xi  = diag\left\{ {{\lambda _1}, \cdots {\lambda _l}} \right\}$. Taking \eqref{64} into \eqref{61} yields ${\bar J_2}{\rm{ = }}\sum\limits_{i = 1}^l {{\lambda _i}} $. This completes the proof.
\end{proof}

Maximizing $tr\left( {\Theta _s^{ - 1}{\Delta _\alpha }} \right)$ improves the replay attack detectability directly when $\chi ^2$ test statistic is adopted. It is of interests to check whether the optimal $M_s$ works for the $LR$ based detection scheme, where the result is given in the following Corollary.

\textbf{Corollary 1}. \emph{Let $M_s \in \mathcal{R}^{{l \times {l}}}$ and
\begin{align}
{M_s} = {\left[ {\begin{array}{*{20}{c}}
{M_{s,1}^T}& \cdots &{M_{s,l}^T}
\end{array}} \right]^T},\label{c1}
\end{align}
where $M_{s,i}^T$ is the generalized eigenvector corresponding to the $i$-th largest generalized eigenvalue for $\left( {{T_{\Delta_\alpha  }}+T_{\Theta_s},{T_{{\Theta _s}}}} \right)$ when $\frac{{\det \left( {{T_{\Delta_\alpha  }}+T_{\Theta_s}} \right)}}{{\det \left( {{T_{{\Theta _s}}}} \right)}} > 1$ and for $\left( {{T_{{\Theta _s}}},{T_{\Delta_\alpha  }}+T_{\Theta_s}} \right)$ when $\frac{{\det \left( {T_{\Delta_\alpha  }}+T_{\Theta_s} \right)}}{{\det \left( {{T_{{\Theta _s}}}} \right)}} \le 1$, then $M_s$ in \eqref{c1} solves the maximization problem $\mathop {\max }\limits_{{M_s}} J_{2,M}$, where
\begin{align}
J_{2,M}=D({f_{\Theta _s^\alpha }}({r_s}(k)),{f_{{\Theta _s}}}({r_s}(k)))\label{c2}.
\end{align}}

\begin{proof}
Based on \eqref{50} and \eqref{c2}, we have
\begin{align}
\mathop {\max }\limits_{{M_s}} J_{2,M}= \frac{1}{2}\mathop {\max }\limits_{{M_s}} \left( {\ln \frac{{\det {\Theta _s}}}{{\det \Theta _s^\alpha }} + tr\left( {\Theta _s^{ - 1}\Theta _s^\alpha } \right)} \right).\label{65}
\end{align}
Let ${d_\Theta } = \frac{{\det \Theta _s^\alpha }}{{\det {\Theta _s}}}$ and it is known that ${d_\Theta } > 0$.
Referring to \eqref{63}, the maximization problem in \eqref{65} is equal to
\begin{displaymath}
\mathop {\max }\limits_{{M_s}} \left( {\ln \frac{1}{{{d_\Theta }}} + {d_\Theta }} \right).
\end{displaymath}
Besides, it is known that $sign\left\{ {{{\partial \left( {\ln \frac{1}{{{d_\Theta }}}  + {d_\Theta }} \right)} \mathord{\left/
{\vphantom {{\partial \left( {\ln \frac{1}{{{d_\Theta }}} - l + {d_\Theta }} \right)} {\partial {d_\Theta }}}} \right.
\kern-\nulldelimiterspace} {\partial {d_\Theta }}}} \right\} = sign({d_\Theta } - 1)$. Since $M_s$ is a square matrix and based on the proof of Theorem 4, $M_s$ in \eqref{c1} implies a maximization of $J_{2,M}$ when $\frac{{\det \left( {{T_{\Delta_\alpha  }}+T_{\Theta_s}} \right)}}{{\det \left( {{T_{{\Theta _s}}}} \right)}} > 1$; when $\frac{{T_{\Delta_\alpha  }}+T_{\Theta_s}}{{\det \left( {{T_{{\Theta _s}}}} \right)}} < 1$, $d_{\Theta}-1<0$, then the matrix pair for determining $M_s$ to maximize $J_{2,M}$ turns to be $\left( {{T_{{\Theta _s}}},{T_{\Delta_\alpha  }}+T_{\Theta_s}} \right)$. This completes the proof.
\end{proof}



\emph{Cumulative sum ratio optimization}. The index $J_1$ focuses on all the detection steps. To concentrate on specific finite step detection when $0<\alpha \le s$, the following index is proposed
\begin{align}
{J_3} = \sum\limits_{\alpha {\rm{ = }}1}^s {\frac{{{M_s}{T_{\Delta ,\Sigma }}M_s^T}}{{{M_s}{T_{{\Theta _s}}}M_s^T}}} .\label{66}
\end{align}
Different from the index $J_1$, the quadratic form in \eqref{66} quantifies the overall influence of the covariances. This ratio index is well adopted to compromise the detection robustness and sensitivity \cite{23_ding2008}. Besides, to enhance the replay attack detection performance for all the possible steps like index $J_2$, the accumulation ratio index is adopted.

Based on \eqref{66}, we have
\begin{align}
{M_s}{T_{\Delta ,\Sigma }}M_s^T = {J_3}{M_s}{T_{{\Theta _s}}}M_s^T.\label{67}
\end{align}
Thus,  ${J_3}$ and $M_s^T$ are the largest generalized eigenvalue and its associated eigenvector of the generalized eigenvalue-decomposition of \eqref{67}, respectively. The solution regarding index $J_3$ is summarized below.

\textbf{Theorem 5}. \emph{Let $M_s^T$ and $\lambda$ be the largest generalized eigenvector and eigenvalue for tuples $\left( {{T_{\Delta ,\Sigma }},{T_{{\Theta _s}}}} \right)$, then $M_s^T$ solves the maximization problem $\mathop {\max }\limits_{{M_s} \in {R^{1 \times {n_z}}}} {J_3}$ with maximum ${\bar J_3}{\rm{ = }}\lambda $.}

\textbf{Remark 9}. \emph{Regarding the case $l>1$, the ratio index $J_3$ can be modified to be a trace ratio problem given by
\begin{displaymath}
{J_{3,M}} = \sum\limits_{\alpha {\rm{ = }}1}^s {\frac{{tr\left( {{M_s}{T_{\Delta ,\Sigma }}M_s^T} \right)}}{{tr\left( {{M_s}{T_{{\Theta _s}}}M_s^T} \right)}}} .
\end{displaymath}
The direct solution to $\mathop {\max }\limits_{{M_s}} {J_{3,M}}$ involves iteration for numerical computation \cite{32_jia2009}. Besides, the solution in \eqref{62} is an approximation of the solution to $\mathop {\max }\limits_{{M_s}} {J_{3,M}}$ when $l>1$ \cite{33_ngo2012}.}

\emph{Detection rate optimization}. Lastly, we focus on the case when false alarm rate $\gamma$ is given, where
\begin{align}
P\left( {r_s^T(k){r_s}(k) > 1|\alpha  = 0} \right) \le \gamma. \label{68}
\end{align}
The threshold is fixed to be $1$ for convenience and can be shifted with respect to $M_s$ under identical detection performance. Based on the Markov inequality, the inequality in \eqref{68} holds when
\begin{align}
\mathcal{E}\{ r_s^T(k){r_s}(k)\}  = tr\left( {{M_s}{T_{{\Theta _s}}}M_s^T} \right) \le \gamma.\label{69}
\end{align}
At the same time, we manage to maximize the detection performance, where trace index
\begin{align}
{J_4} = \sum\limits_{\alpha {\rm{ = }}1}^s {tr\left( {{M_s}{T_{{\Delta _\alpha }}}M_s^T} \right)} \label{70}
\end{align}
is used to quantify the effect of replay attack. The optimization problem is stated by
\begin{align}
&\mathop {\max }\limits_{{M_s}} {J_4},\label{71}\\
&s.t.~~~~ tr\left( {{M_s}{T_{{\Theta _s}}}M_s^T} \right) \le \gamma.\nonumber
\end{align}
The Lagrangian is given by
\begin{displaymath}
L({M_s},\phi ) = \sum\limits_{\alpha {\rm{ = }}1}^s {tr\left\{ {{M_s}{T_{{\Delta _\alpha }}}M_s^T} \right\}}  + \phi \left( {\gamma  - tr\left( {{M_s}{T_{{\Theta _s}}}M_s^T} \right)} \right),
\end{displaymath}
where $\phi$ is the Lagrangian multiplier. Taking derivative of $L({M_s},\phi )$ with respect to $M_s$ yields
\begin{displaymath}
\begin{array}{l}
\begin{aligned}
{M_s}{T_{\Delta ,\Sigma }}& - \phi {M_s}{T_{{\Theta _s}}} = 0,\\
& \Rightarrow {T_{\Delta ,\Sigma }}M_s^T = \phi {T_{{\Theta _s}}}M_s^T.
\end{aligned}
\end{array}
\end{displaymath}
Again we have the generalized eigenvalue-eigenvector problem as in \eqref{67}. With $l>1$, any row of $M_s$ can be determined by the tuples $\left( {{T_{\Delta ,\Sigma }},{T_{{\Theta _s}}}} \right)$. We summarized the solution regarding $J_4$ below.

\textbf{Theorem 6}. \emph{Let $p$ and $\lambda$ be the largest generalized eigenvector and its eigenvalue for $\left( {{T_{\Delta ,\Sigma }},{T_{{\Theta _s}}}} \right)$, respectively, then
\begin{align}
{M_s} = diag\left\{ {{a_1},...{a_l}} \right\}\left[ {\begin{array}{*{20}{c}}
{{p^T}}\\
 \vdots \\
{{p^T}}
\end{array}} \right]\label{72}
\end{align}
solves the optimization problem in \eqref{71} with maximum ${J_4} = \lambda \gamma $, where $a_i \ne 0$.}

\textbf{Remark 10}. \emph{A special case for \eqref{71} is stated as $\mathop {\max }\limits_{{M_s}} \left\{ {{\upsilon _\Delta }\sum\limits_{\alpha {\rm{ = }}1}^s {tr\left\{ {{M_s}{T_{{\Delta _\alpha }}}M_s^T} \right\}}  - {\upsilon _\Theta }tr\left( {{M_s}{T_{{\Theta _s}}}M_s^T} \right)} \right\}$, where ${\upsilon _\Delta },{\upsilon _\Theta } > 0$, and the solution to this special case is covered by the result in Theorem 6. Besides, if matrix 2-norm is adopted, i.e.,
\begin{displaymath}
\mathop {\max }\limits_{{M_s},\left\| {{M_s}} \right\| = 1} \left\{ \begin{array}{l}
{\upsilon _\Delta }{\sigma _i}\left\{ {{M_s}\left[ {\begin{array}{*{20}{c}}
{T_{{\Delta _1},P}^{\frac{1}{2}}}& \cdots &{T_{{\Delta _s},P}^{\frac{1}{2}}}
\end{array}} \right]} \right\}\\
 - {\upsilon _\Theta }\left\| {{M_s}T_{{\Theta _s}}^{\frac{1}{2}}} \right\|
\end{array} \right\}
\end{displaymath}
or $\mathop {\max }\limits_{{M_s}} \left\{ {{\upsilon _\Delta }\left\| {{M_s}\sum\limits_{\alpha {\rm{ = }}1}^s {T_{{\Delta _\alpha },P}^{\frac{1}{2}}} } \right\| - {\upsilon _\Theta }\left\| {{M_s}T_{{\Theta _s}}^{\frac{1}{2}}} \right\|} \right\}$, the solution is identical to the solution in Theorem 6 as well, where ${T_{{\Theta _s}}}$ matters and the solution is optimal to every step. The proof is similar to \cite{23_ding2008} (chapter 7), and thus it is omitted here. Instead of $2$-norm and matrix trace, the $F$-norm may be adopted for optimization, i.e., $\mathop {\max }\limits_{{M_s}} {\left\| {{M_s}{T_{\Delta ,\Sigma }}M_s^T} \right\|_F}$. The solution to the $F$-norm index covers the individual step case (${\left\| {{M_s}{T_{\Delta ,\Sigma }}M_s^T} \right\|_F} \le \sum\limits_{\alpha {\rm{ = }}1}^s {{{\left\| {{M_s}{T_{{\Delta _\alpha }}}M_s^T} \right\|}_F}} $) \cite{34_shang2021}.}

\subsubsection{Unified Solution}
The optimization regarding $J_2$, $J_3$, and $J_4$ aims for the promising time interval for detection. If a specific step is considered, e.g., $\alpha=1$ (related to detection delay) or $\alpha  = {int} \left( {\frac{{s + 1}}{2}} \right)$ (related to the maximum of significance), one can modified ${T_{\Delta ,\Sigma }}$ correspondingly to get the optimal weights.

The relationship between the norm-based and the trace-based indices, i.e., $J_1$ and $J_3$, is to be discussed. If we parameterize the solution to $\mathop {\max }\limits_{{M_s}} {J_3}$ as ${M_s} = \varpi {\Lambda ^{ - 1}}{U^T}$ where $\Lambda$ and $U$ are obtained from \eqref{56}, then \eqref{67} is given by
\begin{align}
\varpi {\Lambda ^{ - 1}}{U^T}{T_{\Delta ,\Sigma }}U{\Lambda ^{ - 1}}{\varpi ^T} = {J_3}\varpi {\varpi ^T}.\label{73}
\end{align}
The equation in \eqref{73} implies the eigenvalue and eigenvector solution to the optimization problem $\max\limits_{{M_s}} {J_3}$, where ${J_{3,g}}$ and ${\varpi ^T}$ are the maximal eigenvalue and its associated eigenvector of ${\Lambda ^{ - 1}}{U^T}{T_{\Delta ,\Sigma }}U{\Lambda ^{ - 1}}$, respectively. Since the solution to the (generalized) eigenvalue-eigenvector problem is not unique, it is easy to found that the optimal solutions to $\max\limits_{{M_s}} {J_i}$ ($i=1,2,3,4$) always owns the form ${M_s} = \varpi {\Lambda ^{ - 1}}{U^T}$. This is the same as the sensitive versus robustness optimization solutions for fault diagnosis \cite{23_ding2008}. Thus, an approximated/unified optimal solution for $M_s$ regarding all the  $J_i$s is ${M_s} = {\Lambda ^{ - 1}}{U^T}$, which only depends on ${T_{{\Theta _s}}}$. Moreover, based on the solution discussion in \cite{35_yu2020} between the LR and the symmetric and positive-definite matrix detection methods, the symmetric and positive-definite matrix method can also be covered here by the unified solution.

\subsubsection{Implementation}
The optimal solutions to $J_2$, $J_3$, and $J_4$ requires to known ${T_{\Delta ,\Sigma }}$, where $P_{x}$, $P_{x,\hat x}$, $P_{wx,\alpha}$, $P_{vx,\alpha}$, $P_{w \hat x,\alpha}$, $P_{v \hat x,\alpha}$, and/or $P_{\hat x}$ are needed. Since the closed-loop system is stable under either static or dynamic controller, the required covariance information can be obtained.

Consider system \eqref{1} running with an observer-based state feedback controller:
\begin{displaymath}
u(k) = K\hat x(k),
\end{displaymath}
where
\begin{displaymath}
\hat x(k + 1) = A\hat x(k) + {B_u}u(k) + L(y(k) - C\hat x(k)).
\end{displaymath}

Define $\bar x(k) = {\left[ {\begin{array}{*{20}{c}}
{{x^T}(k)}&{{{\hat x}^T}(k)}
\end{array}} \right]^T}$, then the closed-loop dynamics for system \eqref{1} is obtained as
\begin{equation}\label{74}
\left\{ \begin{array}{l}
\bar x(k + 1) = \bar A\bar x(k) + {{\bar B}_w}w(k) + {{\bar B}_v}v(k)\\
y(k) = \bar C\bar x + v(k)
\end{array} \right.,
\end{equation}
where
\begin{displaymath}
\begin{array}{l}
\bar A = \left[ {\begin{array}{*{20}{c}}
A&{{B_u}K}\\
{LC}&{A + {B_u}K - LC}
\end{array}} \right],{{\bar B}_w} = \left[ {\begin{array}{*{20}{c}}
{{B_w}}\\
0
\end{array}} \right],{{\bar B}_v} = \left[ {\begin{array}{*{20}{c}}
0\\
L
\end{array}} \right],\\
\bar C = \left[ {\begin{array}{*{20}{c}}
C&0
\end{array}} \right].
\end{array}
\end{displaymath}

Choose $K$ and $L$ appropriately such that \eqref{74} is asymptotically stable. Then the following Lyapunov equation holds:
\begin{equation}\label{75}
\bar A{P_{\bar x}}{\bar A^T} - {P_{\bar x}} + {\bar B_w}Q \bar B_w^T + {\bar B_v}R \bar B_v^T = 0,
\end{equation}
and the solution to \eqref{75} is known as
\begin{equation}\label{76}
{P_{\bar x}} = \sum\limits_{k = 0}^\infty  {{{\bar A}^k}({\bar B_w}Q \bar B_w^T + {\bar B_v}R \bar B_v^T){{({{\bar A}^T})}^k}}.
\end{equation}

From $P_{\bar x}$ one can get $P_x$, $P_{x,\hat x}$ and $P_{\hat x}$ directly. Also, based on the state solution for \eqref{74}, $P_{wx,\alpha}$, $P_{vx,\alpha}$, $P_{w \hat x,\alpha}$, and $P_{v \hat x, \alpha}$ can be obtained.

Based on \eqref{74} and \eqref{76}, it is clear that the closed-loop stability or control performance affects the detectability of replay attack.

If $P_{\bar x}$ is not available, the optimal weighting $M_s$ to $J_1$ or the unified solution is recommended. Alternatively, one can get ${T_{{\Delta _\alpha }}}$ by learning. The procedure is given below

\begin{algorithm}
	\renewcommand{\algorithmicrequire}{\textbf{Input:}}
	\renewcommand{\algorithmicensure}{\textbf{Output:}}
	\caption{Data-driven estimation of ${T_{{\Delta _\alpha }}}$}
	\label{alg1}
	\begin{algorithmic}[1]
		\STATE Determine $N_s$ and set $M_s=I$;
        \STATE Replay system output and/or input data off-line for $2s$ steps for $n_t$ batches, and calculate the residual $r_s^\alpha ({k_i})$ for $0 \le \alpha  \le s$ based on $r_s^\alpha (k) = {Z_s}\left( {{Y_{d,s}}(k) - {H_{u,s}}{U_{d,s}}(k)} \right)$;\\
        \STATE Calculate the approximation of ${\hat T_{{\Delta _\alpha }}} = \frac{1}{{{n_t} - 1}}\sum\limits_{i = 1}^{{n_t}} {r_s^\alpha ({k_i}){{\left( {r_s^\alpha ({k_i})} \right)}^T}}  - {\Theta _s}$.
        \ENSURE  ${\hat T_{{\Delta _\alpha }}}$, where $0<\alpha \le s$.
	\end{algorithmic}
\end{algorithm}

\textbf{Remark 11}. \emph{Based on Algorithm 1, a pure data-driven replay attack detection based on parity space method can be achieved by identifying $N_s$ and learning ${T_{{\Delta _\alpha }}}$ as well as ${\Theta _s}$ \cite{27_ding2014}.}

\subsection{Active Design}
From Section III, it is known that the closed-loop control performance or stability affects the correlations between different variables in the loop, and these correlations really matter the detectability of the replay attack. However, since the control performance is fixed by the closed-loop system dynamics as given in \eqref{74}, it is usually not feasible to redesign $\bar A$ or the closed-loop control performance to trade for a better detection performance against replay attack, if the original detection performance is not satisfying.

As long as the control loop is closed, the system dynamics or $\Delta_{\alpha}$ consisting of controller and plant are fixed. However, when the control system is opened, we know that with or without any input data replay, the system open-loop dynamics affect $\Delta_{\alpha}$ directly. The problem is how and when to open the control loop for replay attack detection. Note that replay attack opens the control loop in stage-$2$. This is the property to be exploited, where an ``attack event-triggered" active attack detection scheme can be designed.

In addition to replay attack detection, the output data-transmission is essential for feedback control. Thus, the expected active detection scheme should keep the quality of feedback information when attack free. In this case, we try to improve the replay attack detectability by redesign the open-loop stability of the attacked system, and the closed-loop control performance remains when attack free.

Referring to the residual generation based on parity space method, it is known that unstable open-loop system usually implies good detection performance regarding replay attack, e.g., $tr\left( {{\Delta _{\alpha ,P}}} \right)$ is relatively large and the indefinite matrices ${P_{{\Delta _\alpha }}}$ and ${P_{u,{\Delta _\alpha }}}$ have more and larger positive eigenvalues \cite{24_zhang2005,25_ferrari2019}. Based on the above analysis, a marginally stable filter method is proposed to compromise the stability and control performance requirement.

The main idea is to cascade a marginally stable filter to the system output. The filter is given by
\begin{equation}\label{77}
\zeta (k + 1) = {A_\zeta }\zeta (k) + {B_\zeta }y(k + 1),
\end{equation}
where $\zeta (k)\in \mathcal{R}^{p}$ is the filter state, and ${A_\zeta }$ and ${B_\zeta }$ are the filter matrices to be determined (${B_\zeta }$ is of full column rank). Instead of $y(k)$, $\zeta (k)$ will be sent to the monitoring and control side. Compared to the passive scheme that no plant-side design is needed, a smart sensor is required here to realize \eqref{77}.

For control purpose, $y(k)$ will be recovered from the received $\zeta (k)$, and the received $\zeta (k)$ will be used for attack detection. The detailed communication scheme is given in Algorithm 2

\begin{algorithm}
	\renewcommand{\algorithmicrequire}{\textbf{Input:}}
	\renewcommand{\algorithmicensure}{\textbf{Output:}}
	\caption{Measurement Information Transmission}
	\label{alg1}
	\begin{algorithmic}[1]
		\STATE For $k=0$, send $\zeta (k) =0$ and $ y(0)$ to the control and monitoring side, and $y(0)$ can be obtained directly;
        \STATE For $k \ge 1$, send $\zeta (k) = {A_\zeta }\zeta (k - 1) + {B_\zeta }y(k)$ to the control and monitoring side, and $y(k) = B_\zeta ^ + \left( {\zeta (k) - {A_\zeta }\zeta (k - 1)} \right)$.
        \ENSURE  $y(k)$.
	\end{algorithmic}
\end{algorithm}

The transformation for $y(k)$ can be implemented conveniently, and the original controller works without any reconfiguration when attack free, because the same controller input $y(k)$ is provided.

Define $\bar x(k) = {\left[ {\begin{array}{*{20}{c}}
{{x^T}(k)}&{{\zeta ^T}(k)}
\end{array}} \right]^T}$. Combining system dynamics and the filter dynamics, the overall dynamics for monitoring is given by
\begin{equation}\label{78}
\left\{ \begin{array}{l}
\begin{aligned}
\bar x(k + 1) = &\underbrace {\left[ {\begin{array}{*{20}{c}}
A&0\\
{{B_\zeta }CA}&{{A_\zeta }}
\end{array}} \right]}_{\bar A}\bar x(k) + \underbrace {\left[ {\begin{array}{*{20}{c}}
{{B_u}}\\
{{B_\zeta }C{B_u}}
\end{array}} \right]}_{\bar B}u(k)\\
 &+ \underbrace {\left[ {\begin{array}{*{20}{c}}
{{B_w}}&0\\
{{B_\zeta }C{B_w}}&{{B_\zeta }}
\end{array}} \right]}_{{{\bar B}_w}}\underbrace {\left[ {\begin{array}{*{20}{c}}
{w(k)}\\
{v(k + 1)}
\end{array}} \right]}_{\bar w(k)},\end{aligned}\\
\bar y(k) = \zeta (k) = \underbrace {\left[ {\begin{array}{*{20}{c}}
0&I
\end{array}} \right]}_{\bar C}\bar x(k).
\end{array} \right.
\end{equation}

Now, based on \eqref{78}, the parity space based replay attack detection can be established as introduced in Sections III and IV.A.

A simple choice for ${B_\zeta }$ and ${A_\zeta }$ is ${B_\zeta }=I$ and ${\sigma _i}({A_\zeta }A_\zeta ^T) = 1$, where $i=1,...,p$. The performance can be checked based on the detectability analysis. For the realization of \eqref{77}, $\left| {{\sigma _i}({A_\zeta })} \right|$ are suggested to be a little smaller than $1$ to compromise the realization error and detection performance.

{A more general case for the marginally stable filter is
\begin{align}
\zeta (k + 1) = {A_\zeta }\zeta (k) + {B_\zeta }y(k).\label{79}
\end{align}
Correspondingly, if the pairs $\left( {\bar A,\bar B} \right)$ and $\left( {\bar A,\bar C} \right)$ based on \eqref{79} are detectable and stabilizable, respectively, the recovery step for $y(k)$ in Algorithm 2 can be omitted, but instead a controller reconfiguration is needed. Thus, the marginally filter in \eqref{77} as well as Algorithm 2 provides a simple and feasible realization for constructing the active method for replay attack detection.

\textbf{Remark 12}. \emph{In the literature, the ``moving target" method is used for attack detection, where an augmented system is introduced as well \cite{16_griffioen2021}. However, the marginally stable filter method proposed here is different from the ``moving target" method. First, we do not assume noises for the auxiliary system. Second, no extra actuator and sensors are needed. Third, a cascade connection between the original system and the marginally stable filter is adopted here which is feasible in practice and no specific original system state and augmented system state interconnection is assumed. Comparing with the moving target method, much less investment/implementation resources is needed for the proposed active design method to detect replay attack, letting alone the proposed passive design method.}

\textbf{Remark 13}. \emph{The control system dynamic property
is exploited for replay attack detection
in this study. The intrinsic security of the control system is established, where the proposed detection scheme does not depend on real-time secret key synchronization and protection.}


\begin{figure}
\begin{center}
\epsfig{file=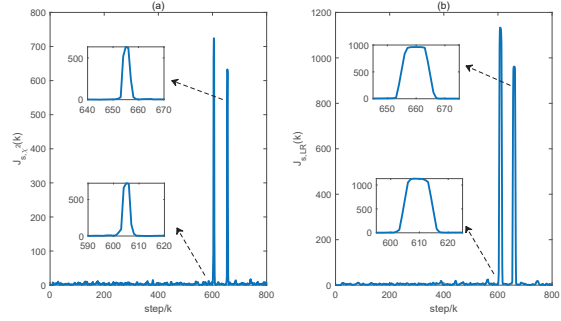,width=7.41846cm, height=4.4625cm}
\caption{$J_{s,\chi ^2}(k)$ and $J_{s,LR}(k)$ under replay attack $y(k)=y(k-300)$ and $u(k)=u(k-300)$ when $k \in [601,650]$.}
\label{fig2}
\end{center}
\end{figure}

\begin{figure}
\begin{center}
\epsfig{file=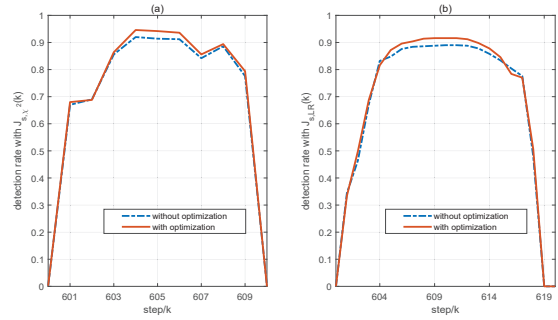,width=7.47558cm, height=4.43394cm}
\caption{Detection rate of replay attack for system \eqref{80} with $y(k)=y(k-300)$ and $u(k)=u(k-300)$ when $k>600$. (a) $J_{s,\chi ^2}(k)$ and (b) $J_{s,LR}(k)$.}
\label{fig3}
\end{center}
\end{figure}

\section{SIMULATION STUDY}
In this section, both passive design and active design examples for replay attack detection are presented.
\subsection{Passive Design Example}
Consider the following system:
\begin{equation}\label{80}
\left\{ \begin{array}{l}
A = \left[ {\begin{array}{*{20}{c}}
{0.9}&0&0\\
{13.4679}&{0.9}&0\\
0&{0.1813}&1
\end{array}} \right],\\
{B_u} = {B_w} = \left[ {\begin{array}{*{20}{c}}
{0.2835}\\
0\\
0
\end{array}} \right],\\
C = \left[ {\begin{array}{*{20}{c}}
0&0&1
\end{array}} \right],~Q = R = 0.01.
\end{array} \right.
\end{equation}
Kalman filter based feedback control is designed to keep the closed-loop control performance.
Both $\chi ^2$ and GLR detection schemes are simulated. Set $s=9$ and implement replay attack $y(k) = y(k - 300)$ and $u(k) = u(k - 300)$ when $k \in [601,650]$.

In Fig. 2, the test statistic responses are illustrated, where it can be found that
\begin{enumerate}
  \item both the occurrence and disappearance of the replay attack can be detected,
  \item and the anomaly response lengths for $\chi ^2$ and GLR schemes are exactly 9 and 18 steps, respectively, which meet the theoretical result.
\end{enumerate}

To show the detection performance further, replay attack is set to be $y(k)=y(k-300)$ and $u(k)=u(k-300)$ when $k>600$. The thresholds for $\chi ^2$ and GLR methods are 20 and 40 respectively. The cases with and without the unified solution of the parity matrix are all simulated 500 times. The detection rate is shown in Fig. 3, where the aimed optimization of the parity matrix improves the detection performance under different test statistics. Also, Fig. 3 shows that the influence of replay attack on the adopted test statistics is detection data length dependent, which provides us a way to distinguish system fault and replay attack without extra investment.

To check the distinction performance between system fault and replay attack further, four fault cases are considered:
\begin{itemize}
  \item constant actuator fault: $B_f=B_u$, $D_f=0$, and $f(k)=1$;
  \item random actuator fault: $B_f=B_u$, $D_f=0$, and $f(k)\sim N(0,0.25)$;
  \item time-varying process fault: $B_f=[1~ 0 ~1]^T$, $D_f=0$, and $f(k)=0.5sin(k/20)$;
  \item random sensor fault: $B_f=0$, $D_f=1$, and $f(k)\sim N(0,0.25)$.
\end{itemize}
Set the faulty time interval to be $(200,600)$. The test statistic responses are illustrated in Fig. 4. Based on Figs. 2 and 4, it can be found that the effects on test statistic driven by system fault and replay attack are different, which implies a distinction between system fault and replay attack.

\begin{figure}
\begin{center}
\epsfig{file=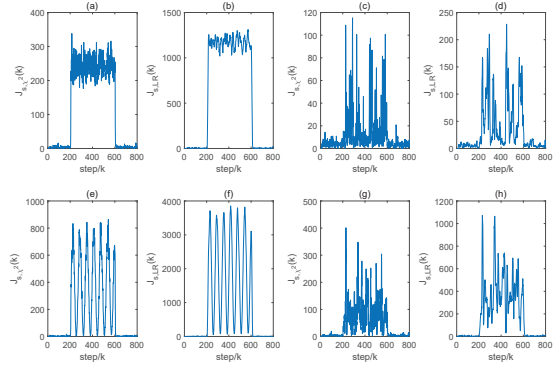,width=7.5564cm, height=5.2884cm}
\caption{$J_{s,\chi ^2}(k)$ and $J_{s,LR}(k)$ under system fault $f(k) \ne 0$ when $k \in (200,600)$. (a) and (b): constant actuator fault with $B_f=B_u$, $D_f=0$, and $f(k)=1$; (c) and (d): random actuator fault with $B_f=B_u$, $D_f=0$, and $f(k)\sim N(0,0.25)$; (e) and (f): time-varying process fault with $B_f=[1~ 0 ~1]^T$, $D_f=0$, and $f(k)=0.5sin(k/20)$; and (g) and (h): random sensor fault with $B_f=0$, $D_f=1$, and $f(k)\sim N(0,0.25)$.}
\label{fig4}
\end{center}
\end{figure}

\subsection{Active Design Example}
It is known from Section III that the detection performance of replay attack based on parity space method is largely determined by the intrinsic dynamics of the closed-loop control system. The detection performance for the control system based on passive design is generally satisfactory; however, poor detection performance cases do exist due to their intrinsic control system dynamics. For example, the control system considered in \cite{6_mo2009}:
\begin{equation}\label{81}
\left\{ \begin{array}{l}
A = {B_u} = {B_w} = C{\rm{ = }}1,\\
Q = 1,~R = 0.1.
\end{array} \right.
\end{equation}
LQG controller is employed for ensuring the closed-loop control performance.

Following the same simulation setting for Fig. 3, both $\chi ^2$ and GLR detection schemes are simulated, with $s=9$, $y(k)=y(k-300)$ and $u(k)=u(k-300)$ when $k>600$. The thresholds for $\chi ^2$ and GLR methods are 20 and 40, respectively. The cases with and without the unified solution of the parity matrix are all simulated 500 times, and the detection rates are illustrated in Fig. 5.

Again, it can be found that the passive design does improve the replay attack detection rate with different test statistics; however, the overall detection performance is still poor.

Based on Algorithm 2, let ${A_\zeta } = {B_\zeta } = 1$, then the proposed active design for replay attack detection is applied. The rest simulation setting is the same as for Fig. 5, and the detection rate is illustrated in Fig. 6.

It can be found from Fig. 6 that the active design improves the detection rate for replay attack largely, even if the original control system dynamics correspond to a poor detection rate as shown in Fig. 5. The effectiveness of the active design is verified.

\begin{figure}
\begin{center}
\epsfig{file=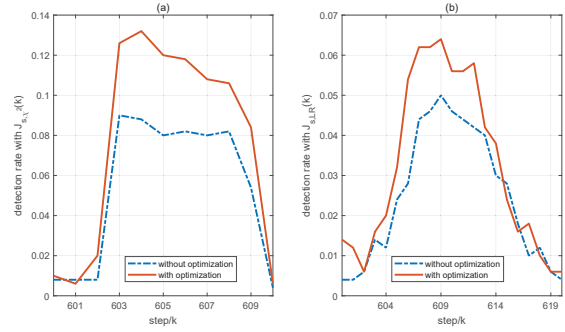,width=7.49343cm, height=4.59102cm}
\caption{Detection rate of replay attack for system \eqref{81} with and without optimization. (a) $J_{s,\chi ^2}(k)$ and (b) $J_{s,LR}(k)$.}
\label{fig5}
\end{center}
\end{figure}

\begin{figure}
\begin{center}
\epsfig{file=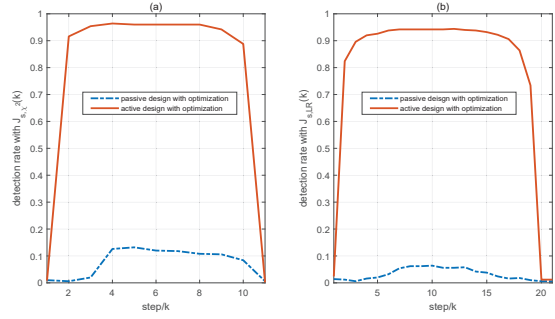,width=7.45054cm, height=4.40186cm}
\caption{Detection rate of replay attack for system \eqref{81}: passive and active design. (a) $J_{s,\chi ^2}(k)$ and (b) $J_{s,LR}(k)$.}
\label{fig6}
\end{center}
\end{figure}


\section{CONCLUSIONS}

In this study, the replay attack detection problem is studied. A new idea is given based on the parity space method. It is proved that the parity space method can catch the replay attack effect on the residual without any further plant side modification. The replay attack effect on residual is quantified, followed by the detection performance analysis based on $\chi ^2$ and LR test statistics, respectively. To improve the detection performance with the parity space based residual generation, both passive and active detection design are proposed, where the basic idea is to enlarge the replay attack effect on residual comparing to the attack-free case. By exploiting the specific property of the obtained replay attack effect, four optimization problems are formulated to obtain the optimal parity matrices as the passive detection design, and a marginally stable filter method is proposed as the active detection design to enlarge the replay attack effect for detection further.

Different from the reported methods, the proposed method can deal with input and output data replay attack, isolate system fault and replay attack, maintain certain control performance, and be implemented conveniently and efficiently. The quantitative effect analysis of replay attack on residual indicates the security as well as vulnerability level of the control system under replay attack. The given detection performance can be checked based on control system model or data information. Thus, data-driven design of parity space method for replay attack detection can be further studied from both the defense and attack perspectives.

\begin{appendix}
The derivation of ${cov} (h({r_s}(k)))$. Based on \eqref{44}, we have
\begin{align}
cov& (h({r_s}(k))) \nonumber\\
=~& {cov} \Big(\ln \frac{{\det {\Theta _s}}}{{\det \Theta _s^\alpha }} + r_s^T(k)\Theta _s^{ - 1}{r_s}(k)\nonumber\\
 &- r_s^T(k){(\Theta _s^\alpha )^{ - 1}}{r_s}(k)\Big)\nonumber\\
 =~& {cov} \left( {r_s^T(k)\Theta _s^{ - 1}{r_s}(k)} \right) + {cov} \left( {r_s^T(k){{(\Theta _s^\alpha )}^{ - 1}}{r_s}(k)} \right)\nonumber\\
 &-2{cov} \left( {r_s^T(k)\Theta _s^{ - 1}{r_s}(k),r_s^T(k){{(\Theta _s^\alpha )}^{ - 1}}{r_s}(k)} \right).\label{82}
\end{align}
As $\mathcal{E}(r_s(k))=0$ and $\mathcal{E}(r_s^{\alpha}(k))=0$, based on \cite{36_rencher2008} (Theorem 5.2c), it is known that
\begin{align}
{cov} \left( {r_s^T(k)\Theta _s^{ - 1}{r_s}(k)} \right) = 2tr\{ {(\Theta _s^{ - 1}\Theta _s^\alpha )^2}\},\label{83}
\end{align}
\begin{align}
{cov} \left( {r_s^T(k){{(\Theta _s^\alpha )}^{ - 1}}{r_s}(k)} \right) = 2l. \label{84}
\end{align}
Based on \cite{37_liu2009} (Lemma 2.2), we have
\begin{align}
&{cov} \left( {r_s^T(k)\Theta _s^{ - 1}{r_s}(k),r_s^T(k){{(\Theta _s^\alpha )}^{ - 1}}{r_s}(k)} \right)\nonumber\\
 &= 2tr\{ \Theta _s^{ - 1}\Theta _s^\alpha \}.\label{85}
\end{align}
Taking \eqref{83}, \eqref{84}, and \eqref{85} into \eqref{82} yields
\begin{displaymath}
{cov} (h({r_s}(k))) = 2tr\{ {(\Theta _s^{ - 1}\Theta _s^\alpha )^2}\}  + 2l - 4tr\{ \Theta _s^{ - 1}\Theta _s^\alpha \} .
\end{displaymath}
\end{appendix}


%
%


\bibliographystyle{IEEEtran}
\bibliography{replay_attack_detection_arXiv}

\end{document}